\documentclass[sigconf,nonacm]{acmart}
\setcopyright{none}
\usepackage{multirow}
\usepackage{changepage}

\usepackage{graphics}
\usepackage{amsmath}	   \usepackage{shadow}        \usepackage{hhline}        \usepackage{tabularx}      \usepackage{latexsym}      \usepackage{fancyhdr}      \usepackage{verbatim}
\usepackage{xspace}		   \usepackage[super]{nth}  \usepackage{color}       \usepackage{balance}     \usepackage{graphicx}
\usepackage{float}
\usepackage{caption}
\usepackage{makecell}

\usepackage{subcaption}

\usepackage{shortcuts}
\usepackage[absolute,showboxes]{textpos}

\renewcommand{\paragraph}{\vspace{3pt}\noindent\textbf}
\begin{document}

\date{}

\setlength{\TPHorizModule}{\paperwidth}
\setlength{\TPVertModule}{\paperheight}
\begin{textblock}{0.8}(0.1,0.02)
     \noindent
     \small
     \textcolor{blue!80!black}{If you cite this paper, please use the CCS'23 reference:
     Gibran Gomez, Kevin van Liebergen, and Juan Caballero.
     Cybercrime Bitcoin Revenue Estimations: Quantifying the Impact of Methodology and Coverage.
     In \textit{CCS '23: Proceedings of the 2023 ACM SIGSAC Conference on Computer and Communications Security, 2023, Pages 3183–3197}.
     DOI: \url{https://doi.org/10.1145/3576915.3623094}.}
\end{textblock}

\thispagestyle{empty}

\title[Cybercrime Bitcoin Revenue Estimations: Quantifying the Impact of Methodology and Coverage]{Cybercrime Bitcoin Revenue Estimations:\\Quantifying the Impact of Methodology and Coverage}

\author{Gibran Gomez}
\affiliation{
  \institution{IMDEA Software Institute}
  \institution{Universidad Polit\'{e}cnica de Madrid}
  \country{Madrid, Spain}
}
\email{gibran.gomez@imdea.org}

\author{Kevin van Liebergen}
\affiliation{
  \institution{IMDEA Software Institute}
  \institution{Universidad Polit\'{e}cnica de Madrid}
  \country{Madrid, Spain}
}
\email{kevin.liebergen@imdea.org}

\author{Juan Caballero}
\affiliation{
  \institution{IMDEA Software Institute}
  \country{Madrid, Spain}
}
\email{juan.caballero@imdea.org}

\pagenumbering{arabic}

\begin{abstract}
\noindent 

Multiple works have leveraged the public Bitcoin ledger 
to estimate the revenue cybercriminals obtain from their victims.
Estimations focusing on the same target often do not agree,
due to the use of different methodologies,
seed addresses, and time periods.
These factors make it challenging
to understand the impact of their methodological differences.
Furthermore, they underestimate the revenue due to the 
(lack of) coverage on the target's payment addresses, 
but how large this impact remains unknown.

In this work, we perform the first systematic analysis on
the estimation of cybercrime bitcoin revenue.
We implement a tool that can replicate the different
estimation methodologies.
Using our tool we can quantify, in a controlled setting, 
the impact of the different methodology steps.
In contrast to what is widely believed,
we show that the revenue is not always underestimated.
There exist methodologies that can introduce huge overestimation.
We collect \allAddr payment addresses and use them to
compare the financial impact of 6 cybercrimes
(ransomware, clippers, sextortion, Ponzi schemes, giveaway scams,
exchange scams)
and of \numGroups cybercriminal groups. 
We observe that the popular multi-input clustering
fails to discover addresses for 40\% of groups.
We quantify, for the first time, the impact of the 
(lack of) coverage on the estimation.
For this, we propose two techniques to achieve high coverage, 
possibly nearly complete, on the DeadBolt server ransomware.
Our expanded coverage enables estimating DeadBolt's revenue 
at \$2.47M, 39 times higher than the estimation using 
two popular Internet scan engines. 

\end{abstract}




\maketitle
\section{Introduction}
\label{sec:intro}

The Bitcoin ecosystem has attracted cybercriminal activities such as
ransomware~\cite{bitiodineSpagnoulo,behindLiao,economicConti,cerber,trackingHuang,ransomwareClouston,pony},
thefts~\cite{fistfulMeiklejohn},
scams~\cite{spamsPaquetClouston,ponziBartoletti,fistfulMeiklejohn,xia2020characterizing,bartoletti2021cryptocurrency,li2023giveaway},
human trafficking~\cite{backpagePortnoff},
clippers~\cite{wyb},
cryptojacking~\cite{huang2014botcoin,tekiner2021sok},
hidden marketplaces~\cite{travelingChristin,cybercriminalLee,dreadRon}, and
money laundering~\cite{inquiryMoser}.
Cybercriminals often request payments in bitcoins from
victims that fall for their scams or are infected with their malware.
The public nature of the Bitcoin ledger has been leveraged 
by multiple works
to estimate the revenue cybercriminals obtain 
from victims~\cite{huang2014botcoin,bitiodineSpagnoulo,behindLiao,economicConti,trackingHuang,ponziBartoletti,cybercriminalLee,ransomwareClouston,spamsPaquetClouston,xia2020characterizing,oosthoek2022tale,li2023giveaway}.
An accurate estimation of the financial impact on victims 
is fundamental for understanding the cybercrime ecosystem, 
e.g., for comparing the revenue of different types of cybercrime,
such as ransomware versus sextortion. 
It is also critical for triaging,  
i.e., assigning adequate investigative resources to each cybercriminal 
group based on their impact.
Underestimating the financial impact of a cybercriminal group 
may convey the wrong impression that it is unimportant,
removing resources from its analysis and defense
and thus allowing its operations to continue unfettered.
On the other hand, overestimating the financial impact may wrongly 
identify small players as dominant,
assigning them unnecessary investigative resources.

Starting from a set of \emph{seed} Bitcoin addresses, 
known to belong to the same cybercriminal group 
(e.g., ransomware family, Ponzi scheme) or 
to the same type of cybercrime (e.g., ransomware, sextortion),
estimation works apply different methodologies to quantify the 
financial impact on the victims.
Some works simply add deposits to the input seed addresses, 
while others apply a combination of expansions
to discover additional payment addresses 
(e.g., multi-input 
clustering~\cite{bitcoin,evaluatingAndroulaki,quantitativeRon,fistfulMeiklejohn}
and change address heuristics~\cite{fistfulMeiklejohn,evaluatingAndroulaki,cookieGoldfeder,automaticErmilov,kappos22peel}) 
and filters to remove unrelated addresses and deposits.
Estimations focusing on the same target often do not agree,
due to the use of different methodologies,
seed addresses, and time periods (i.e., block heights).
Furthermore, these factors make it very challenging 
to understand to what degree each methodology step is responsible for 
differences in the estimation.

In this work, we perform the first systematic analysis on 
the estimation of cybercrime bitcoin revenue. 
We quantify the impact in the estimation of the methodology used
and the limited seed coverage.
In detail, 
we survey prior estimation works providing 
a detailed analysis of their methodologies and 
concrete takeaways.
We implement a tool that can replicate the different methodologies. 
We apply our tool to estimate the same target using \numMethods methodologies, 
while fixing the set of seeds and the blockchain height. 
This allows us to quantify how differences in the methodology 
affect the estimation.
In contrast to what is widely believed, 
we show that estimations do not always underestimate the revenue.
There exist estimation methodologies 
that can introduce huge overestimation
such as those that do not filter 
seeds that are online wallets in exchanges
and those using change address heuristics.
We collect \allAddr cybercrime Bitcoin payment addresses 
from publicly available sources.
We use this dataset to compare,
using a consistent methodology, the impact of 6 cybercrimes
(ransomware, clippers, sextortion, Ponzi schemes, giveaway scams, 
exchange scams)
and of \numGroups groups running Ponzi schemes, ransomware, and clippers 
(which replace addresses copied into the clipboard with their own).
We observe that the top cybercrime groups by revenue are dominated by 
ransomware and the most successful ones operate in the 
ransomware-as-a-service model~\cite{raas}, 
although there are also Ponzi schemes and clipper families 
for which we observe more than one million USD revenue. 
There are also 14 groups whose estimated revenue is below \$20, 
likely indicating limited coverage on their payment addresses.
Our results highlight the limited effectiveness of existing expansions.
The popular multi-input (MI) clustering 
fails to discover additional addresses for 40\% of 
the \numGroups groups,
likely indicating cybercriminals are actively evading it,
while change address expansion introduces large false positives. 

An intrinsic issue in cybercrime revenue estimations 
is that they underestimate 
the real revenue 
because they start from a limited set of seeds 
(oftentimes only one) while the cybercriminals may use a large number 
of addresses to receive victim payments.
To this day, no work has quantified the impact of the (lack of) coverage on 
the estimation.
This requires a vantage point that allows to observe all victim payments.
We perform the first quantification of this issue.
For this, we focus on the DeadBolt server ransomware family, 
which infects network-attached storage (NAS) devices~\cite{deadbolt}. 
We start by collecting 4,997 DeadBolt payment addresses from two 
Internet scan engines~\cite{shodan,censys}.
We estimate DeadBolt's conversion rate from infections to payments to 
be 0.7\%. A unique characteristic of DeadBolt is that it releases the decryption key 
on the blockchain upon receiving the victim's payment.
We propose two novel techniques, 
leveraging unique characteristics of DeadBolt's key release transactions,
to obtain very high coverage, possibly nearly complete, on the payments
DeadBolt receives from victims.
Using the 34 seeds with victim payments collected from the scan engines
(which MI clustering cannot expand)
we would have estimated a very modest revenue of 2.826 BTCs or \$63K. 
By applying our coverage-expanding techniques, 
we instead estimate 98.350 BTC, 35 times higher, 
and our USD estimation is \$2.47M, 39 times higher.
The vantage point provided by the scan engines only identifies 
2.6\% of victim payments due to issues such as scan frequency or 
infections happening before the scanning begins.
Still, Internet scan engines arguably provide higher coverage on 
server ransomware (i.e., 34 DeadBolt seeds) that is typically available 
for other groups such as desktop ransomware 
(i.e., a median of one seed per group). 
Thus, for other groups, the coverage impact may be even larger.

Our coverage results critically indicate that
even if a family or campaign is estimated to have low revenue,
it could still have a significant non-measured financial impact on victims.
Thus, estimations should also consider other impact metrics beyond the revenue,
e.g., the number of family samples observed.

\vspace{+0.2cm}
\noindent This work provides the following main contributions:

\begin{itemize}

\item We perform the first systematic analysis of 
cybercrime bitcoin revenue estimations. 
We build a tool that implements the different estimation methodologies 
and use it to quantify the impact of each methodology step 
in the estimation.
We show that some methodologies can produce huge overestimation.

\item We quantify, for the first time, the impact of the 
(lack of) coverage on the estimation. 
For this, we propose two novel techniques to achieve 
high, possibly complete, coverage of the victim payments received by 
the DeadBolt server ransomware.
The USD revenue DeadBolt collects from victims is 39 times larger
than what would have been estimated by collecting seeds from 
two popular Internet scan engines.

\item We compare the bitcoin revenue obtained by 6 cybercrimes 
and \numGroups cybercriminal groups.
Cybercriminals may be actively evading the popular multi-input clustering, 
which does not discover additional addresses for 40\% of groups.

\item We have released our estimation tool and DeadBolt dataset~\cite{wybrepo}.

\end{itemize}

\begin{table*}[t]
\centering
\scriptsize
\begin{tabular}{|l|c|l|l|r|r|r|r|c|c|c|c|c|c|c|c|c|c|l|}
\cline{6-17}
\multicolumn{5}{c|}{} &
\multicolumn{5}{c|}{{\bf Seeds}} & 
\multicolumn{3}{c|}{{\bf Expand}} &
\multicolumn{4}{c|}{{\bf Filtering}} &
\multicolumn{2}{c}{} \\
\hline
\textbf{Work} &
\textbf{Year} &
\textbf{Crime} &
\textbf{Platform} &
\textbf{Blk Height} &
\crotate{Seeds} & 
\crotate{Labels} & 
\crotate{Micro Payments} & 
\crotate{Seed Clustering} & 
\crotate{Seeds Available} & 
\crotate{Multi-input (MI)} & 
\crotate{Change Address (CA)} & 
\crotate{Exploration} & 
\crotate{Value Filtering (VF)} & 
\crotate{Time Filtering (TF)} & 
\crotate{Online Wallets (OW)} & 
\crotate{Double-Counting (DC)} & 
\crotate{Payment day rate} &
\textbf{Methodology} \\
\hline
Huang et al.~\cite{huang2014botcoin} & 2014 & Cryptojacking & - & 2013-11-30 & 290 & 10 & \N & \N & \N & \Y & \N & \N & \N & \N & \N & \N & \Y & DD+MI \\
Spagnuolo et al.~\cite{bitiodineSpagnoulo} & 2014 & Ransomware & BitIodine & 2013-12-15 & $\ge$12 & 1 & \N & \N & \N & \Y & \Y & \N & \Y & \N & \N & \N & \N & DD+MI+CA-VF \\
Liao et al.~\cite{behindLiao} & 2016 & Ransomware & - & 2014-01-31 & 2 & 1 & \N & \N & \cite{behindLiao} & \Y & \Y & \N & \Y & \Y & \N & \N & \Y & DD+MI+CA-VF-TF \\
Conti et al.~\cite{economicConti} & 2018 & Ransomware & - & 2017-12-07 & 128 & 20 & \N & \N & \cite{economicContiDataset} & \Y & \Y & \N & \Y & \Y & \N & \N & \Y & DD+MI+CA-VF-TF \\
Huang et al.~\cite{trackingHuang} & 2018 & Ransomware & BlockSci & 2017-08-31 & 25 & 10 & \Y & \N & \N & \Y & \N & \lc & \Y & \N & \lc & \N & \Y & DD-OW+MI-VF \\
Bartoletti et al.~\cite{ponziBartoletti} & 2018 & Ponzi & - & - & 32 & 32 & \N & \N & \cite{ponziBartolettiDataset} & \Y & \N & \N & \N & \N & \N & \N & \Y & DD+MI \\
Lee et al.~\cite{cybercriminalLee} & 2019 & Dark Web & BlockSci & 2018-04-30 & 85 & - & \N & \N & \N & \Y & \Y & \N & \N & \N & \lc & \N & \Y & DD-OW+MI+CA \\
Paquet-Clouston et al.~\cite{ransomwareClouston} & 2019 & Ransomware & GraphSense & 489,181 & 7,118 & 35 & \N & \N & \cite{ransomwareCloustonDataset} & \Y & \N & \N & \N & \Y & \N & \lc & \Y & DD+MI-TF-DC \\
Paquet-Clouston et al.~\cite{spamsPaquetClouston} & 2019 & Sextortion & GraphSense & 573,989 & 245 & - & \N & \Y & \cite{spamsCloustonDataset} & \Y & \N & \N & \Y & \N & \lc & \lc & \Y & DD-OW+MI-VF-DC \\ Xia et al.~\cite{xia2020characterizing} & 2020 & Exchange scams & - & 2019-09-23 & 66 & - & \N & \Y & \cite{xia2020characterizingDataset} & \N & \N & \N & \N & \N & \N & \N & \N & DD \\
Oosthoek et al.~\cite{oosthoek2022tale} & 2022 & Ransomware & - & - & 7,321 & 87 & \N & \N & \cite{ransomwhere} & \N & \N & \N & \N & \N & \N & \N & \Y & DD \\
Li et al.~\cite{li2023giveaway} & 2023 & Giveaway scams & - & 2022-07-01 & 860 & - & \N & \Y & \cite{li2023giveawayDataset} & \N & \N & \N & \N & \N & \N & \N & \N & DD \\
\hline
\end{tabular}
\caption{Related work on estimating cybercrime Bitcoin financial impact.
A tick (\Y) indicates a property is implemented by the work, 
a cross (\N) no support,
a half-filled circle (\lc) partial support, and
a dash (-) that the paper does not specify it.
}
\label{tab:related}
\vspace{-0.3cm}
\end{table*}

\section{Bitcoin Revenue Estimation}
\label{sec:estimation}

This section surveys prior works 
that estimate cybercrime bitcoin 
\emph{revenue}~\cite{huang2014botcoin,bitiodineSpagnoulo,behindLiao,economicConti,trackingHuang,ponziBartoletti,cybercriminalLee,ransomwareClouston,spamsPaquetClouston,xia2020characterizing,oosthoek2022tale,li2023giveaway}.
These works do not estimate the \emph{profit} the cybercriminals made, 
as that would require subtracting from the revenue 
the unknown expenses they incurred~\cite{thomas2015framing}.
Estimations may target  
a specific cybercriminal group 
(e.g., malware family, Ponzi scheme, scam campaign) or 
a (type of) cybercrime 
(e.g., ransomware)
by aggregating revenue of multiple groups in the same cybercrime 
(e.g., multiple ransomware families). 

At a high level, all the works follow the same approach. 
They take as input a set of \emph{seed} addresses 
known to receive victim payments.
They optionally expand the seed addresses 
to obtain an \emph{expanded set} of addresses that belong to the same owners 
as the seeds.
If no expansions are used, the expanded set only contains the seeds.
Then, they obtain all the deposits to addresses in the expanded set.
Next, they optionally apply filtering to remove unrelated 
addresses and deposits.
This process results in an estimation of BTC revenue.
Finally, the BTC amount is converted to fiat currency (typically US Dollars). 
While the general process is shared, there exist methodological 
differences at each of the above steps that can lead to 
differing estimations for the same input seeds.

Table~\ref{tab:related} summarizes the \numpapers surveyed works.
To select these works, we first examined papers published since 2009
(the initial Bitcoin protocol release year)
in top computer security venues.
To identify other articles published in smaller venues and
pre-print repositories, we analyzed the references of those initial works.
Additionally, we searched on engines like Google Scholar for
combinations of keywords, including
\textit{bitcoin}, \textit{payments},
\textit{ransomware}, and \textit{scam}.
We limit the selection to papers published on peer-reviewed venues
(and technical reports with public datasets)
that include an estimation of cybercriminal bitcoin revenue. 
We exclude works that analyze Bitcoin abuse by cybercriminals
but do not include a financial estimation
(e.g.,~\cite{tekiner2021sok,wyb}),
those performing estimations on other cryptocurrencies 
(e.g., Monero~\cite{hong2018you,bijmans2019just}), and
those estimating revenue using transaction data not 
from the Bitcoin blockchain (e.g.,~\cite{laarschot2021risky}).
The rest of this section details the above steps using the different 
parts of Table~\ref{tab:related}.

\subsection{Seeds}
\label{sec:seeds}

We call \emph{payment addresses} to Bitcoin addresses 
where victims are requested to send their payments.
Payment addresses may be obtained 
from social media (e.g.,~\cite{bitiodineSpagnoulo}),
threat intelligence reports by security companies (e.g.,~\cite{economicConti}), 
by running malware samples on a sandbox (e.g.,~\cite{trackingHuang}), 
from scam emails (e.g.,~\cite{spamsPaquetClouston}), 
from Tor hidden services (e.g.,~\cite{cybercriminalLee}), and
from scam websites (e.g.,~\cite{li2023giveaway}).
Payment addresses can be split into those that have received some deposit,
which are called \emph{seeds},  
and those that have not (yet). 
This determination, and generally the whole estimation, 
needs to be performed at some particular height of the Bitcoin blockchain. 
Unfortunately, only two works provide the specific block height 
used~\cite{ransomwareClouston,spamsPaquetClouston}.
Most works mention a day,
but there are roughly 144 blocks in a day, as
a new block is minted every 10 minutes.

Cybercriminals may or may not reuse payment addresses across victims. 
In the extreme, they could use a single payment address for all victims, 
or generate a different address for each victim. 
Oftentimes, a middle ground is used with a 
(potentially large) pool of payment addresses 
that are reused with some (potentially low) probability.
Payment addresses with deposits are the input seeds to the estimation.
The number of seeds 
ranges from only one up to 7,036 addresses.
However, the high number of seeds in two 
works~\cite{ransomwareClouston,oosthoek2022tale}
is due to the outlier Locky ransomware with over 7K addresses.
Overall, the median number of seeds per group is 1
(measured on the dataset in Section~\ref{sec:cybercrimes}).
Seeds are the critical starting point for any estimation.
To enable replicability, authors should release the seeds used,
which happens for 8 of the \numpapers works.

One challenge when each victim is assigned a unique payment address 
is that only a small fraction of payment addresses receives 
payments and thus can be used as seeds. 
For example, in their sextortion paper, 
Paquet Clouston et al.~\cite{spamsPaquetClouston} extracted 
12,533 payment addresses, but only 2\% had deposits. Furthermore, for some address collection methods like 
running malware on a sandbox or collecting emails marked as spam 
(prior to the user receiving them), if the address is unique for 
each victim then none of the collected payment addresses would have payments, 
as there are no real victims. 
To address this situation, Huang et at.~\cite{trackingHuang} 
perform micro-payments to payment addresses.
Micro-payments simulate the payment of a victim, but their value is 
much smaller than the requested payment value. 
Their goal is incentivizing the cybercriminals to move the micro-payment
so that it can reveal other payment addresses through the expansions presented
in Section~\ref{sec:expansion}.
For example, a micro-payment of a few hundred satoshis
is small enough that to move those funds the recipients need to 
combine them with other Unspent Transaction Outputs (UTXOs) in a 
multi-input (MI) transaction so that 
the transaction fee can be paid. 
This enables discovering those additional inputs through 
MI clustering, as described in Section~\ref{sec:expansion}. 
The larger the micro-payment, the larger the incentive for the 
cybercriminals to move the gifted funds. 
However, cybercriminals could ignore any payments not matching the 
expected values.

Seeds may be labeled with the group to which they belong,
which allows to perform separate estimations for each group.
For example, the label may capture the malware family that uses the 
seed, obtained from external reports or 
from the AV labels of the samples~\cite{avclass}. 
But, some collection methods such as 
visiting Tor hidden services~\cite{cybercriminalLee}, 
visiting scam websites~\cite{xia2020characterizing,li2023giveaway}, and
examining spam emails~\cite{spamsPaquetClouston} 
may collect addresses that belong to different groups. 
To perform per-group estimates, 
these works first need to cluster the addresses using external 
information such as the content of the websites or the spam emails.
When seeds are unlabeled and no clustering is performed, 
it is only possible to provide an estimation 
of the type of cybercrime all addresses are associated with
(e.g., the Dark Web~\cite{cybercriminalLee}).
The quality of the labels is fundamental to the estimation. 
If a seed is incorrectly labeled, that will 
inflate the estimation of its group.

Some works extract Bitcoin addresses from Web pages using regular 
expressions~\cite{cybercriminalLee,xia2020characterizing,li2023giveaway}.
Since Bitcoin addresses are hashes, such regular expressions may generate 
false positives. 
It is important to validate the matches by verifying the checksum embedded 
in valid addresses, e.g., using online blockchain explorers or 
by using tools that perform such validation 
during the extraction~\cite{iocsearcher}.

\take{To be replicable, works should release their payment addresses
(as seeds may change with the block height), 
the clustering results (if any), and 
the block height used for the estimation.}

\subsection{Expansions}
\label{sec:expansion}

Given a set of seeds, the simplest estimation consists of adding 
all direct deposits the seeds have received, 
i.e., for each transaction where a seed address appears in an output slot,
accumulate the value of the seed's output slot. 
This simple estimation is used in three surveyed 
works~\cite{xia2020characterizing,oosthoek2022tale,li2023giveaway}.
However, if the campaign uses a large number of payment addresses, 
the seeds may only receive a small
portion of the revenue received from victims, 
i.e., there may exist many unknown payment addresses that also received 
victim payments.
To discover previously unknown payment addresses, 
it is common to apply expansions to the seeds
that identify additional addresses that 
also belong to the seed owners. 
Then, the same estimation as above is performed 
on the expanded set of addresses
(i.e., seeds plus additional addresses the expansions identified). 
Two main expansions are used in the surveyed works: 
\emph{multi-input clustering} used by 8 works and 
\emph{change address} heuristics used by 4 works.
They can be combined,
e.g., all works using change address also use multi-input clustering.
We also discuss an \emph{exploration} expansion,
which generalizes the approach used in one work for one group.

\paragraph{Multi-input clustering.}
The most popular expansion 
is multi-input (MI) 
clustering~\cite{bitcoin,evaluatingAndroulaki,quantitativeRon,fistfulMeiklejohn}.
It assumes that input addresses to the same transaction
have the same owner because their private keys are
used together to sign the transaction.
Clusters can be created transitively.
If a transaction has addresses $A$ and $B$ as inputs, and
another transaction has $B$ and $C$ as inputs, then $A$, $B$ and $C$
are all clustered together and have the same owner.
One exception
is CoinJoin~\cite{coinjoinMaxwell} transactions
where a group of users creates a single transaction
that simultaneously spends all their inputs into a shuffled list of
outputs.
For this reason, before computing MI clustering,
it is common to apply proposed heuristics to identify
CoinJoin transactions~\cite{cookieGoldfeder,blocksci}.
Kappos et al.~\cite{kappos22peel} recently proposed a 
machine learning (ML) classifier to reduce the false negatives of 
the CoinJoin heuristics.
However, their classifier has not yet been used in estimations.

Once the MI clusters are computed, 
the seeds and all addresses in the MI cluster of 
a seed are added to the expanded set.
If the seeds belong to different MI clusters, 
the expanded set contains the union of all addresses in those MI clusters
(including the seeds).
Of the 9 works using MI clustering, 
5 use open-source platforms to compute it: 
one uses BitIodine~\cite{bitiodine}, 
two use BlockSci~\cite{blocksci}, and 
two use GraphSense~\cite{graphsense}. 
The rest use their own implementation. 
As far as we know, no work has compared that different implementations 
of MI clustering indeed produce the same results. 

While MI clustering is an extremely popular and reliable 
expansion~\cite{nick2015msc,harrigan2016unreasonable},
there are two caveats that could affect the estimation. 
First, MI clustering captures \emph{same ownership}. 
If the same cybercriminals run multiple campaigns, 
MI clustering may add to the expanded set addresses 
from campaigns different from the one the seeds belong to.
If the goal is to estimate the revenue of a specific campaign, 
rather than the revenue of a cybercriminal group,
or of a type of cybercrime, 
this may introduce overestimation.

\take{MI clustering may add to the estimation other campaigns 
from the same owner.}

Second, \emph{double ownership} is common in services 
(e.g., exchanges) offering online wallets for their users.
In that scenario, the service 
owns the address
(i.e., has the private key),
but the address is handled (and thus indirectly owned)
by the user for which it is created.
Performing MI clustering on an online wallet address 
can bring into the expanded set thousands (and even millions) of 
addresses that also belong to the service, 
but are unrelated to the estimation as they belong to 
other customers of the service.
For seeds that are online wallets, 
only their direct deposits should be considered, 
i.e., other addresses in their MI cluster 
should not be added to the expanded set.
Failure to filter seeds that are online wallets, 
can make the estimation hugely overestimate the financial impact.
We discuss the filtering of online wallets in Section~\ref{sec:filtering}.

\take{
It is possible for MI clustering to largely 
overshoot the actual revenue
if some seeds are online wallets in services like exchanges, and
the estimation includes all the cluster deposits.
For seeds that are online wallets in services, 
the expanded set should contain only the seeds, i.e.,
it should not contain other addresses in the MI clusters of those seeds.}

\paragraph{Change address.}
Bitcoin's UTXO-based model does not allow the partial spending of
transaction outputs.
Since the sum of input values to a transaction may be larger than the 
amount that needs to be paid, a \emph{change address} can be used 
by the owner of the input addresses to collect back the change. 
Several heuristics have been proposed for identifying which output slot 
in a transaction
is the change address~\cite{fistfulMeiklejohn,evaluatingAndroulaki,cookieGoldfeder,automaticErmilov,kappos22peel}.
Once identified, the change address
(and other addresses in its MI cluster)
are added to the expanded set.
Three works (\cite{bitiodineSpagnoulo,behindLiao,economicConti}) 
use the change address heuristic by Androulaki et al., 
which checks transactions with two output addresses. 
If one output address is fresh (i.e., never used before) and the other is not, 
the fresh address is considered the change address.
Instead, Lee et al. use one of the variants implemented by 
BlockSci~\cite{blocksci}, but do not detail which one.
Recently, Kappos et al.~\cite{kappos22peel} compare different
change address heuristics, showing that most have high false positives (FPs).
They propose a new heuristic to reduce FPs, 
which has not yet been used by any estimation.

\take{The change address heuristics currently used by estimation works 
can generate a large number of false positives, 
and thus overestimate the financial impact.} 

\begin{figure}[t]
  \centering
  \includegraphics[width=\linewidth]{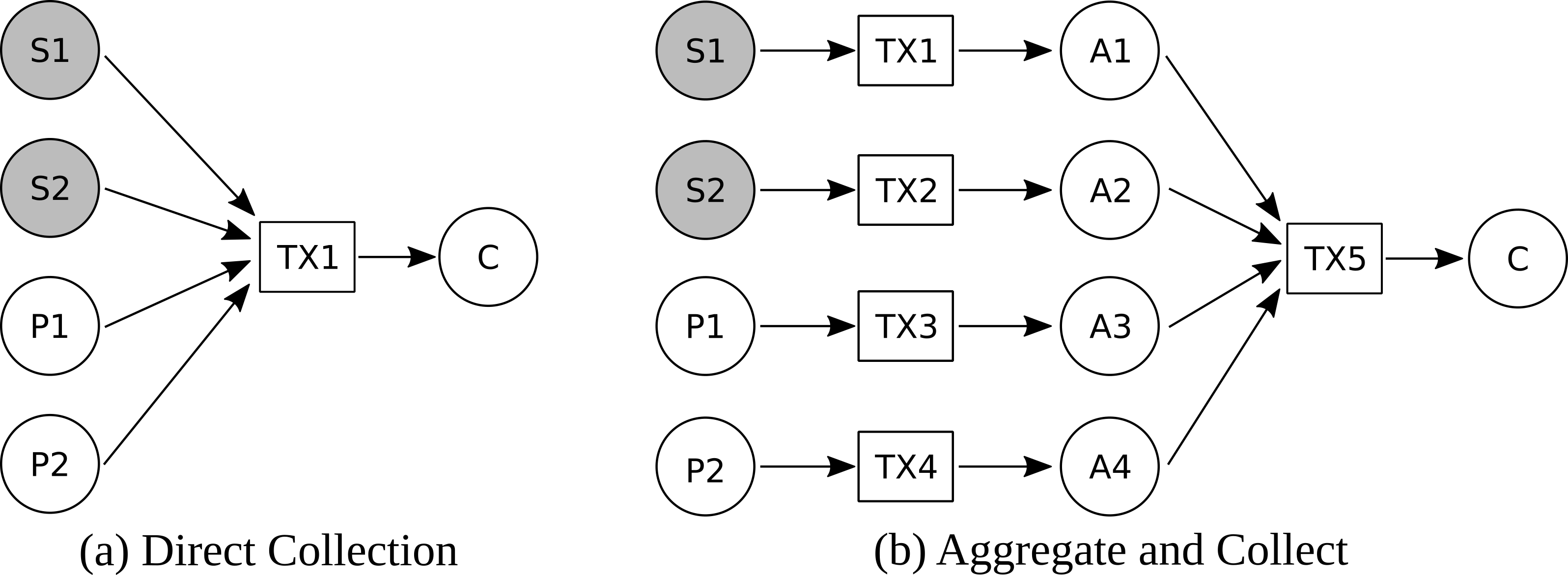}
  \caption{Two approaches to collect payments from 
two seeds $S_1, S_2$ (in gray) and two unknown payment address $P_1,P_2$ into a 
collector address $C$.
On the left, MI clustering discovers the unknown addresses $P_1,P_2$, 
while on the right it does not.}
  \label{fig:mic}
  \vspace{-12pt}
\end{figure}

\paragraph{Exploration.}
Cybercriminals may accumulate the received payments 
prior to cashing them out or sending them to a mixer to 
obfuscate their origin. 
MI clustering may capture such aggregation if multiple payment addresses are 
used as input to the same transaction.
However, the effectiveness of applying MI clustering on the seeds 
depends on how the cybercriminals withdraw their funds.
Consider the example in Figure~\ref{fig:mic}(a),
where the cybercriminals use two seeds ($S_1,S_2$) and 
two unknown payment addresses ($P_1,P_2$) as input to a 
withdrawal transaction (withdrawal for short) 
that accumulates the payments into a \emph{collector} address $C$.
In this case, MI clustering would identify the previously unknown 
payment addresses ($P_1,P_2$).
Instead, in Figure~\ref{fig:mic}(b) the funds are first moved 
from the 4 payment addresses into 4 aggregation addresses 
($A_1-A_4$) in separate transactions, and then collected using a MI
transaction into collector $C$. 
In this case, MI clustering on the two seeds would not identify the 
unknown payment addresses or the aggregators.

Withdrawals from the seeds
evade MI clustering if they use a single input address.
We can classify withdrawals based on their number of
distinct input and output addresses.
We focus on distinct addresses
rather than on transactions slots
because a transaction can use multiple input slots
for the same address.
For example, a payment address may be reused for multiple victims;
each victim payment generates a different UTXO and all UTXOs are
consumed by the same withdrawal.
We say a transaction is \emph{1-to-n} ($n\ge1$) if it has a single
input address, regardless of its number of input slots.
In Section~\ref{sec:evasion} we measure that 40\% of groups 
only use MI-defeating 1-to-n withdrawals from their seeds, 
indicating that this evasion is widely used.

Figure~\ref{fig:mic}(b) captures how the Cerber ransomware operated.
To handle this family, Huang et al. proposed to 
manually add the aggregators ($A_1,A_2$) into the set of seeds. 
That allows MI to discover the other two aggregators ($A_3,A_4$) and 
thus perform a more accurate estimation. 
In Section~\ref{sec:discussion} we discuss how we believe their approach 
could be generalized to address MI clustering evasion.

\take{Cybercriminals can use 1-to-n withdrawals from 
payment addresses to defeat MI clustering.}

\subsection{Filtering}
\label{sec:filtering}

Not all deposits to the expanded set are necessarily victim payments.
The seeds and the additional addresses the expansions identify 
could be used for other purposes, 
e.g., for aggregating funds. 
For these reasons, 8 of the 12 works apply filters to exclude 
deposits that do not look like victim payments.
We describe them next.

\paragraph{Value filtering.}
If the amounts the victims should pay are known, 
deposits for other amounts can be excluded.
One consideration is that some victims
may ignore that, in addition to the requested amount, 
they also need to pay a transaction fee. 
This results in some victim payments not reaching the requested amount 
since the transaction fee is discounted prior to depositing the funds.
Some payments can also be slightly higher than the requested amount
because the victim conservatively increases the amount to make sure 
any fees are covered.
To account for these small deviations in amounts, some works apply 
an epsilon around the known payment values~\cite{behindLiao,economicConti}.
A limitation of value filtering is that, due to limited coverage, 
it may not be possible to know all valid amounts
(e.g., each victim could be given a different amount).
Thus, some victim payments could be incorrectly excluded.
For example, 13 of the 20 ransomware families 
considered in~\cite{economicConti} have no seeds matching the expected 
ranges, and thus cannot be estimated. 
Another limitation is that value filtering does not apply to cases where 
the victim decides how much to pay, 
e.g., giveaway scams~\cite{li2023giveaway}.

\paragraph{Time filtering.}
If the time periods when a campaign was active are known, 
deposits to the expanded set outside of those periods can be excluded.
This helps exclude other uses of the payment addresses before or after 
a campaign happened.
Similar to value filtering, due to limited coverage, 
it may not be possible to know all periods when a campaign was active,
which may exclude valid victim payments.

Some works combine value and time filtering by 
taking as input a list of time ranges, 
each associated with a list of values victims could pay on the period.
This handles campaigns where the amounts change over time, 
e.g., they are lowered when the BTC conversion rate spikes to 
avoid exorbitant fees that discourage victims to pay. 

\take{If payment addresses are not reused, 
value and time filtering are needless.
The problem of coverage affects not only the discovery of seeds, 
but also value and time filtering ranges, 
possibly introducing underestimation.}

\paragraph{Online wallet filtering.}
If a seed is an online wallet in a service, 
only direct deposits to that seed should be considered.
Other addresses in its cluster should not be included in the expanded set.
Two works~\cite{spamsPaquetClouston,cybercriminalLee} 
filter exchange clusters by looking for clusters that are outliers, 
in terms of large number of addresses, 
large total amount received, and
cluster age. 
Unfortunately, these filters are not detailed. 
In addition, two works identify exchange clusters using tag databases that 
associate addresses with additional information like their owner. 
Lee et al.~\cite{cybercriminalLee} use the public 
WalletExplorer~\cite{walletexplorer} tag database and 
Huang et al.~\cite{trackingHuang} a proprietary database 
from Chainalysis~\cite{chainalysis}.
However, Huang et al.~\cite{trackingHuang} 
do not use the tags to identify seeds that are online wallets.
Instead, they filter deposits to the expanded set that do not 
originate from exchanges,
since most victims likely do not own BTCs and need to purchase them from 
exchanges. 
However, they discarded this filter due to concerns about the coverage of the 
tag database.
To address this intrinsic limitation of tag databases,
Gomez et al.~\cite{wyb} recently proposed to complement tag databases 
with an ML classifier to identify (untagged) exchanges. 
However, their exchange classifier has not been used in any estimation.

\take{The identification of exchange clusters is better performed using 
a combination of tag databases and a machine learning classifier.}

\paragraph{Double-counting.}
Simply adding all deposits to addresses in the expanded set,
without considering their procedence, may lead to double-counting, 
i.e., counting the same payment multiple times,
thus overestimating the revenue.
The simplest such double-counting can occur
when, in a transaction, one input address also appears in the outputs, 
e.g., an input address is used as a change address.
While this is not recommended for privacy reasons, it often happens in practice.
Imagine Alice has two BTCs in one UTXO from a previous payment
and wants to send one BTC to Bob.
Instead of creating a fresh change address, 
Alice uses the sender address as the change address.
After the payment, the balance of Alice's address will be one BTC,
but the total amount deposited to her address will be three BTCs,
two BTCs from the original payment that created the UTXO
plus one BTC she received as change from this transaction.

More generally, double-counting can occur when a transaction that 
deposits into an address in the expanded set has an input 
address that also belongs to the expanded set.
Imagine two deposits. 
In the first one, address \emph{A} which is not in the expanded set 
deposits some funds into address \emph{B} which is part of the expanded set. 
In the second one, address \emph{B} moves the received funds 
to address \emph{C} which is also in the expanded set.
The funds \emph{B} received from \emph{A} are first counted. 
Then, if we add the second deposit that only moves funds between addresses 
inside the expanded set, we are double counting the funds 
that \emph{B} received from \emph{A} as also being received by \emph{C}.

Only the two works by 
Paquet-Clouston et al.~\cite{ransomwareClouston,spamsPaquetClouston} 
try to avoid such double-counting by filtering transactions. 
In~\cite{ransomwareClouston}, they define collectors 
as addresses receiving deposits 
from at least two addresses in the expanded set. 
Collectors belonging to the expanded set were filtered to avoid double-counting.
However, it is possible to double count
with a single deposit if the change address is one of the inputs. 
In~\cite{spamsPaquetClouston} they 
exclude transactions with at least one input address and one output 
address in the expanded set.

One issue to keep in mind 
is that any filter that tries to minimize double-counting 
will be affected by the coverage issue, 
i.e., it will not be able to filter deposits between two MI clusters 
that belong to the group if one of the two MI clusters has not been 
included in the expanded set, i.e., no seed is available from that MI cluster.

\take{To avoid overestimating the revenue, 
deposits into the expanded set that have an input address also belonging 
to the expanded set should be filtered.
Double-counting filtering is also affected by the 
lack of seed coverage.}

\subsection{Conversion Rates}
\label{sec:conversion}

Adding the amounts of the filtered deposits to the expanded set 
produces an estimation of BTC revenue.
Due to Bitcoin's highly volatile conversion rate, 
the estimated BTCs can have very different values over time.
Converting the BTC amount into US Dollars 
makes it easier to understand the revenue and to compare it with other
cybercrime, e.g., those not abusing Bitcoin.
To perform this conversion, 9 of the \numpapers works use the 
conversion rate from BTC to USD on the day the deposit was received.
This conversion rate captures the financial impact on victims, 
i.e., how much it cost the victims to pay assuming they bought the BTCs 
from an exchange just prior to paying.
Two works~\cite{bitiodineSpagnoulo,xia2020characterizing}
use instead the conversion rate on a specific day,
e.g., the day of the analysis.
Since cybercriminals may not cash out victim payments immediately, 
this approach includes the rise or 
depreciation in value of the payments
from the day the payment was received until the day of the conversion rate. 
Finally, one work uses the highest and lowest conversion rates in their
analysis period~\cite{li2023giveaway}, 
providing a USD range.

\take{To estimate the financial impact on victims using fiat currency, 
it is recommended to apply the conversion rate on the day each payment was 
received.}

\begin{table*}
\centering
\scriptsize
\begin{tabular}{l|l|l|r|r|r|r|r|r}
\hline
	\textbf{Work} & \textbf{Blk Height} & \textbf{Methodology} & \textbf{Seeds} & \textbf{Clust.} & \textbf{Addrs.} & \textbf{Payments} & \textbf{BTC} & \textbf{USD} \\
\hline
	Spagnuolo et al.~\cite{bitiodineSpagnoulo}	& 2013-12-15 & DD+MI+CA-VF & $\ge$12 & $\ge$12 & 2,118 & 771 & 1,226.0000 & 1,100,000 \\
					Liao et al.~\cite{behindLiao}			& 2014-01-31 & DD+MI+CA-VF-TF & 2 & 1 & 968 & 795 & 1,128.4000 & 310,472 \\
		Conti et al.~\cite{economicConti}	& 2017-12-07 &  DD+MI+CA-VF-TF & 4 & 1 & 956 & 804 & 1,403.7548 & 449,275 \\
						Huang et al.~\cite{trackingHuang}		& 2017-08-31 & DD-OW+MI-VF & 2 & 2 & 4,457 & - & - & 667,000 \\
						Paquet-Clouston et al.~\cite{ransomwareClouston}	& 489,181 & DD+MI-TF-DC & 2 & 1 & 944 & - & 1,511.7100 & 519,991 \\
					\hline
\end{tabular}
\caption{Previous estimations on CryptoLocker.}
\label{tab:cryptolocker}
\vspace{-0.4cm}
\end{table*}

\section{Methodology Impact}
\label{sec:methodimpact}

This section 
quantifies how the different methodology options
presented in Section~\ref{sec:estimation} 
impact the estimation. 
We first introduce the analyzed methodologies and datasets in 
Section~\ref{sec:methodologies}, 
then quantify their impact on the CryptoLocker ransomware 
in Section~\ref{sec:cryptolocker}, and 
on our whole dataset in Section~\ref{sec:other}.

\subsection{Methodologies and Datasets}
\label{sec:methodologies}

Given the 2 main expansions 
(excluding the exploration as it is only used for one ransomware family 
in one work) 
and 4 filters detailed in Table~\ref{tab:related}, 
64 different methodologies can be used for the estimation. 
We have created a naming scheme to refer to those methodologies.
The name always starts with \emph{DD} which is an abbreviation for 
direct deposits.
Then, it can have at most 6 parts for 
multi-input clustering (\emph{MI}), 
change address expansion (\emph{CA}), 
online wallet filtering (\emph{OW}),
value filtering (\emph{VF}), 
time filtering (\emph{TF}), and
double-counting filtering (\emph{DC}). 
We prefix expansions with a plus sign 
and filters with a minus sign. 
The last column in Table~\ref{tab:related} captures
the methodology used by each work using our naming scheme.

We have implemented an extension to the WYB platform~\cite{wybrepo}, 
itself built on top of BlockSci~\cite{blocksci},
which can produce any of the estimations. 
Given as input a set of seeds, a block height, 
a set of value and time filtering ranges, and 
a methodology string using our naming scheme;
it produces the estimation using that methodology.
While our extension implements all previously used methodologies, 
the results it produces are not guaranteed to be the same as 
those obtained in the works in Table~\ref{tab:related}, 
even when using the same methodology on the same seeds. 
Differences may happen because the implementation of the expansions and filters 
may not exactly match the ones in those works.
For example, we use BlockSci to obtain clusters using 
multi-input and change address expansions, 
while other works use other open-source platforms or their own implementations.

\paragraph{Selected methodologies.}
We analyze \numMethods selected methodologies, 
which cover those used in the surveyed works and additional ones 
illustrating interesting cases. 
The first 5 estimations test the filters without applying expansions. 
We exclude the filter of online wallets as it only applies to expansions.
\emph{DD} is the simplest estimation that sums 
all deposits to the seeds. 
\emph{DD-VF}, \emph{DD-TF}, and \emph{DD-VF-TF} apply 
value filtering, time filtering, and both, respectively, 
on the direct deposits to the seeds 
using the 7 ranges proposed by Conti et al.~\cite{economicConti}.
\emph{DD-DC} removes overlaps in the direct deposits 
caused by transactions 
where seeds appear as both input and output.

The next 6 estimations correspond to expanding using only MI clustering 
and then applying the different filters.
The final 4 estimations correspond to expanding with both MI clustering and 
change address expansions and then applying the different filters.
MI clustering and change address expansions both
use the clusters that BlockSci precomputes 
for the given block height.
MI clustering excludes CoinJoin transactions using the heuristics in BlockSci.
\emph{DD+MI} and \emph{DD+MI+CA} first query BlockSci for the clusters 
of the seeds 
(precomputed using only MI clustering or both expansions, respectively) and 
add the seeds and other addresses in the clusters of the seeds to the 
expanded set.
Then, they obtain all deposits to the expanded set and sum their value
without any filtering.
\emph{DD-OW+MI} and \emph{DD-OW+MI+CA} first query BlockSci 
for the clusters of the seeds.
Then, they check whether the seed clusters appear in the WYB tag database
marked as service clusters.
For any seed cluster identified as a service,
only the seed is added to the expansion set.
For other seed clusters, the seed and
other addresses in the cluster are added to the expanded set.
Then, they obtain all deposits to the expanded set and sum their value
without applying any filtering.
\emph{DD-OW+MI-DC} applies \emph{DD-OW+MI} 
but avoids double-counting by filtering deposits to the expanded set 
that have an input address belonging to any of the clusters 
in the expanded set.
All other estimations correspond to the ones above with 
value filtering, time filtering, or both, 
applied on the deposits to the expanded set. 

\paragraph{Datasets.} 
We collect all publicly available
labeled datasets of cybercrime Bitcoin seeds that we are aware of.
These include all datasets mentioned in Table~\ref{tab:related},
as well as an additional dataset of 9,478 addresses used by
clippers~\cite{wyb}.
Since the ransomware datasets overlap,
we use the Ransomwhere dataset~\cite{ransomwhere},
which contains the seeds used in prior works.
For scam datasets containing addresses for
different blockchains~\cite{xia2020characterizingDataset,li2023giveawayDataset}
we focus exclusively on the Bitcoin addresses.
Other sources of Bitcoin cybercrime seeds exist such as
abuse databases (e.g.,~\cite{bitcoinabuse}),
but they do not provide a reliable categorization of cybercrime type.
The six datasets used,
summarized in Table~\ref{tab:cybercrime},
in total comprise \allAddr
cybercrime Bitcoin addresses, 
of which \allSeeds have deposits at
block height 785,100 (April 12, 2023).
Some works
clustered the seeds,
but their datasets do not include the cluster identifier for each seed.
To estimate individual groups,
we use a subset of \groupSeeds labeled seeds belonging to \numGroups groups:
88 ransomware families,
22 clipper families, and
31 Ponzi schemes.

\subsection{Impact on CryptoLocker}
\label{sec:cryptolocker}

We first analyze the CryptoLocker ransomware 
because its revenue has been estimated in 
five prior works with varying results, 
as illustrated in Table~\ref{tab:cryptolocker}. 
For each prior estimation, the table shows
the maximum block height (or date) considered
in the estimation, 
the methodology used, 
the number of seeds, 
the number of MI clusters for the seeds, 
the total addresses in the expanded set, 
the number of victim payments identified after filtering, and 
the estimation in BTC and USD.
The table illustrates that the estimations can widely vary, 
e.g., the highest estimation of \$1.1M is 3.5 times larger than the
lowest estimation of \$310K.
Due to the estimations use 
different seeds, methodologies, and block heights,
it is complicated to isolate the impact of each methodology decision 
on the estimation. 
To address this issue, we perform estimations using different methodologies, 
but on a fixed set of seeds and on the same block height.
This allows us to analyze the impact of each methodology decision.

Of the 5 prior estimations, 
two do not specify the seed addresses 
used~\cite{bitiodineSpagnoulo,trackingHuang}. 
Two use the same pair of seeds~\cite{behindLiao,ransomwareClouston}, and 
one~\cite{economicConti} uses 4 seeds
(the two in~\cite{behindLiao,ransomwareClouston} and
two additional ones).
We fix the block height at 498,150 (December 7, 2017)
and perform two estimations: 
one starting with the two seeds in~\cite{behindLiao,ransomwareClouston} and
another from the four seeds in~\cite{economicConti}.
We convert from BTC to USD using the rate at the time of 
each payment, obtained from CoinDesk~\cite{coindesk}.

\begin{table}
\centering
\scriptsize
\begin{tabular}{l|r|r|r|r}
\hline
\textbf{Methodology} & \textbf{Addr.} & \textbf{Deposits} & \textbf{BTC} & \textbf{USD} \\ \hline
DD & 2 & 54 & 100.8608 & \$12,532 \\ DD-VF & 2 & 48 & 94.8625 & \$11,786 \\ DD-TF & 2 & 53 & 100.8607 & \$12,531 \\ DD-VF-TF & 2 & 47 & 93.8625 & \$11,667 \\ DD-DC & 2 & 54 & 100.8608 & \$12,532 \\ \hline
DD+MI & 968 & 1,101 & 1,544.9144 & \$528,046 \\ DD+MI-VF-TF & 968 & 803 & 1,130.1121 & \$309,935 \\ DD-OW+MI & 968 & 1,101 & 1,544.9144 & \$528,046 \\ DD-OW+MI-VF-TF & 968 & 803 & 1,130.1121 & \$309,935 \\ DD-OW+MI-DC & 968 & 1,077 & 1,511.7579 & \$520,238 \\ DD-OW+MI-VF-TF-DC & 968 & 801 & 1,110.3121 & \$304,791 \\ \hline
DD+MI+CA & - & - & - & - \\
DD-OW+MI+CA & 2 & 54 & 100.8608 & \$12,532 \\ DD+MI+CA-VF-TF & - & - & - & - \\
DD-OW+MI+CA-VF-TF & 2 & 47 & 93.8625 & \$11,667 \\ \hline
\end{tabular}
\caption{CryptoLocker estimations using different methodologies starting from the two seeds used in~\cite{behindLiao,ransomwareClouston}.
A dash indicates the estimation did not finish in 5 days.
}
\label{tab:estimations2}
\end{table}

\begin{table}
\centering
\resizebox{\columnwidth}{!}{\begin{tabular}{l|r|r|r|r}
\hline
\textbf{Methodology} & \textbf{Addr.} & \textbf{Deposits} & \textbf{BTC} & \textbf{USD} \\
\hline
DD & 4 & 321 & 70,426.0298 & \$11,442,063 \\ DD-VF & 4 & 59 & 123.6568 & \$18,006 \\ DD-TF & 4 & 288 & 57,987.8966 & \$10,094,385 \\ DD-VF-TF & 4 & 50 & 106.8525 & \$15,062 \\ DD-DC & 4 & 227 & 41,614.1073 & \$6,339,388 \\ \hline
DD+MI & 12,770,529 & 27,617,860 & 134,110,936.7365 & \$26,405,734,619 \\ DD+MI-VF-TF & 12,770,529 & 39,015 & 54,091.2840 & \$32,786,920 \\ DD-OW+MI & 970 & 1,368 & 71,870.0835 & \$11,957,577 \\ DD-OW+MI-VF-TF & 970 & 806 & 1,143.1021 & \$313,330 \\ DD-OW+MI-DC & 970 & 1,208 & 34,771.1654 & \$5,453,205 \\ DD-OW+MI-VF-TF-DC & 970 & 797 & 1,112.8791 & \$305,080 \\ \hline
DD+MI+CA & - & - & - & - \\
DD-OW+MI+CA & 4 & 321 & 70,426.0298 & \$11,442,063 \\ DD+MI+CA-VF-TF & - & - & - & - \\
DD-OW+MI+CA-VF-TF & 4 & 50 & 106.8525 & \$15,062 \\ \hline
\end{tabular}
}
\caption{CryptoLocker estimations using different methodologies starting from the four seeds used in~\cite{economicConti}. 
A dash indicates the estimation did not finish in 5 days.
}
\label{tab:estimations4}
\vspace{-0.4cm}
\end{table}

\paragraph{Two seeds.}
Table~\ref{tab:estimations2} presents the estimations 
on the two seeds from~\cite{behindLiao,ransomwareClouston}.
The estimations without expansions are pretty consistent 
in the range of \$11.6K--\$12.5K.
MI clustering is quite successful at discovering additional CryptoLocker 
addresses increasing the estimations 26--42 times 
to \$304.7K--\$528.0K.
Among filters, value filtering has the largest impact, 
especially after MI clustering, 
indicating that some deposits may not correspond to victim payments.
Online wallet filtering has no effect as none of the seeds are online wallets.
The double-counting filter reduces the estimations by 1.5\%--1.7\%. 

Surprisingly, we are unable to compute some estimations with change address 
expansion.
The cluster with the seeds returned by BlockSci has 160M addresses.
Due to the huge number of deposits to that cluster (261M), 
the estimation does not finish in 5 days and we stop it.
Three works in Table~\ref{tab:cryptolocker} claim to use 
change address expansion. 
However, their estimates are very close to ours using only MI clustering.
One possibility is that the change address expansion they implemented was not 
transitive, and thus avoided the explosion affecting BlockSci's implementation.
Similar to what has been observed in prior work~\cite{kappos22peel}, 
our results indicate that change address heuristics can have high 
false positives. 
We further examine this issue for all groups in Section~\ref{sec:other}.
Once online wallets are filtered, the estimations with change address 
expansion finish.
This happens because the huge cluster that contains the seeds also 
contains multiple addresses tagged as exchanges.
Thus, the cluster is considered to belong to an exchange, 
and both seeds are (incorrectly) considered online wallets.
Thus, \emph{DD-OW+MI+CA} matches the results for \emph{DD} and 
\emph{DD-OW+MI+CA-VF-TF} matches the results for \emph{DD-VF-TF}.

\paragraph{Four seeds.}
Table~\ref{tab:estimations4} presents the estimations 
on the four seeds from~\cite{economicConti}.
In this case, the estimations without expansions widely vary 
from \$15K up to \$11.4M, due to the additional seeds receiving
deposits of millions of USD.
Value and time filtering removes most of those deposits as likely 
not being victim payments, resulting in a \$15K estimate, only 30\% 
higher than with two seeds. 
After applying MI clustering, the expanded set grows to 12.7M addresses, and 
the estimation is an astronomical 26 Billion USD.
The online wallet filter uncovers the reason for this
by identifying one of the additional seeds as 
an online wallet in the btc-e exchange and the other as belonging to an
incorrect huge cluster due to an address whose private key is 
the empty string~\cite{wyb}.
Considering both additional seeds as online wallets reduces the estimation 
to \$11.9M, and value and time filtering brings it down to \$313K.
A key question is whether the two additional seeds are indeed related 
to CryptoLocker. 
We are inclined to think they are not, but we have no definite proof. 
Still, this experiment illustrates how failing to consider seeds 
that are online wallets or belong to other problematic MI clusters 
can hugely overestimate the financial impact. 
Value filtering significantly reduces the impact 
of incorrectly applying MI clustering on online wallets, but it cannot
entirely correct the problem as some fraction of the massive amount of 
deposits will match the expected value ranges.

\ignore{
\paragraph{Estimation using different methodologies.}
Two previous works analyze the CryptoLocker operation
beginning from the same two seeds,~\cite{behindLiao}, and~\cite{ransomwareClouston}.
Both seeds are expanded to a single MI cluster of 968 addresses.
The first one~\cite{behindLiao} analyzes deposits to this cluster
up to the last period of activity of Cryptolocker,
on January 31, 2014.
Authors found a total of 1,071 deposits to this cluster,
a total of 1,541.39 BTC.
Next, they use a \vtf
to identify ransom payments among all deposits,
accordingly to seven known CryptoLocker campaigns.
They identify 795 ransom payments,
for a total of 1,128.40 BTC.
This filtering technique drops 25.8\% deposits that are not ransom payments
and reduces in 26.8\% the total BTC revenue.
By implementing the \vtf in~\cite{economicConti}
we find 803 ransom payments to the CryptoLocker cluster,
for a total of 1,130.11 BTC.
The difference of 1.7 BTC with respect to Liao et al. results
comes from the margin amount allowed by Conti et al.,
which is related to the users not considering the transaction fees in the payments.
The margin used by Conti et al.
is based on the transaction fees,
which is more relaxed that the one used by Liao et al.,
fixed in 0.1 or 0.05 BTC.
As we implemented the \vtf of Conti et al.,
we compare our results with the mentioned 803 payments.
By using our \novestimation,
two ransom payments are dropped.
Just 801 deposits are reported as ransom payments
for a total of 1,110.31 BTC.
This means that 19.8 BTC were sent from within the cluster,
and the \vtf was not able to identify them as redundant.

In the second work~\cite{ransomwareClouston}, the CryptoLocker operation is analyzed
up to October 28, 2017,
three months after the last deposit to the cluster,
on July 10, 2017.
Instead of using a \vtf,
authors identify \emph{key addresses}
receiving two or more deposits
from any address within the CryptoLocker cluster.
Authors find 24 of such key addresses already present in the MI cluster
and 26 that are new (e.g., not present in the MI cluster).
Authors remove the 24 key addresses already present in the MI cluster
and estimate using all the direct deposits sent to the 944 addresses left,
for a total of 1,511.71 BTC.
In total, there are 1,101 direct deposits
to the 968 addresses of this cluster,
for an amount of 1,544.91 BTC,
so this reduction of 2.48\% on the number of addresses
lead to a reduction of 2.15\% on the amount of BTC received by the cluster.
There are two main differences with our \novestimation approach:
(a) Authors of~\cite{ransomwareClouston} excluded addresses receiving
two or more payments from addresses within the same cluster (24 \emph{key addresses}),
thus if an address received just one payment from within its cluster,
it won't be excluded, and the payment will count for their estimation,
even if it is a loop deposit from the same address
where the sender sends the change to itself;
(b) we exclude payments instead of excluding addresses,
thus payments received by \emph{key addresses}
coming from outside the cluster will count for our estimation.
By using \novestimation without the \vtf,
we found 1,077 deposits to this cluster,
for a total of 1,511.76 BTC.
The difference is very small because
most of the overlapping payments produced by the cluster
were sent to the set of 24 \emph{key addresses}
identified by Paquet-Clouston et al.

In a third work~\cite{economicConti},
authors add two more seeds to the CryptoLocker operation.
One of these seeds is an online-wallet of BTC-e~\cite{btceTakedown},
a cryptocurrency exchange taken down due to its connections with money laundering.
The other is an online-wallet that belongs to a supercluster
associated to multiple services,
resulting due to the reuse of the Bitcoin address produced from the empty string~\cite{wyb}.
As mentioned before,
online-wallets may introduce a large amount of noise to the estimation.
This is the case for these two wallets,
that most probably are not receiving ransom payments,
but act as collectors used to cashing-out the funds obtained from illicit activities.
Both seeds are expanded to 12.7M addresses,
which receive more than 134M BTC in more than 27,6M deposits.
However, the \novestimation
reveals that just 55.5M BTC come from outside the clusters.
By excluding all unrelated addresses
from the estimation by the tag filter,
the estimation decreases to 267 deposits,
a total of 70,325.17 BTC directly sent to the online-wallet seeds.
However, only 150 deposits come from outside the clusters,
for a total of 34,038.50 BTC.
When applying the \vtf,
only three deposits are found to be ransomware payments,
with a value of 12.99 BTC,
with no overlaps between them.
Interestingly, just one of the two addresses receive such payments,
which indicate that the other seed was not used for ransomware payments.

Finally, when applying the \novestimation over the four seeds,
only 797 deposits are found to be ransom payments.
Seven payments found when individually analyzing
the two seeds of Liao et al. (801 payments)
and to the two online-wallet seeds added by Conti et al. (3 payments)
are dropped,
This means that some transactions that matched the \vtf
were actually sending funds to each other.
Our final estimation without overlap deposits
for the CryptoLocker operation is of 1,112.88 BTC,
which are equivalent to 305,080 USD.
\gibran{Add some comment on CA results}.
}

\subsection{Impact on the Whole Dataset}
\label{sec:other}

This section quantifies the impact of expansions and filters 
on the whole dataset.
We exclude the value and time filters since we only have 
ranges for a few ransomware families.
All experiments are performed on block height 785,100 (April 12, 2023).

\begin{table}
\centering
\scriptsize
\begin{tabular}{l|c|r|r|r|r|r}
\hline
	\textbf{Cybercrime} & \textbf{Dataset} & \textbf{Seeds} & \textbf{Labels} & \textbf{OW} & \textbf{BTC} & \textbf{USD}  \\
\hline
		Ransomware & ~\cite{oosthoek2022tale} & 7,352 & 88 & 12 & 31,434.4069 & \$115,513,569 \\
	Giveaway scams & ~\cite{li2023giveaway} & 494 & - & 17 & 386.0907 & \$12,207,462 \\
	Ponzi schemes & ~\cite{ponziBartoletti} & 32 & 31 & 7 & 7,956.3913 & \$5,029,029 \\
	Clippers & ~\cite{wyb} & 637 & 22 & 23 & 891.5331 & \$2,814,443 \\
	Sextortion & ~\cite{spamsPaquetClouston} & 248 & - & 5 & 305.9584 & \$1,504,437 \\
	Exchange scams & ~\cite{xia2020characterizing} & 53 & - & 2 & 117.2569 & \$1,021,764 \\

\hline
		\multicolumn{2}{l|}{\textbf{Total}} & 8,816 & 141 & 66 & 41,091.6373 & \$138,090,704 \\
\hline
\end{tabular}
\caption{Cybercrime type estimates on April 12, 2023 
using the seeds in public datasets and a \emph{DD-OW+MI-DC} estimation.}
\label{tab:cybercrime}
\vspace{-0.5cm}
\end{table}

\paragraph{Online wallet filter impact.}
We use BlockSci to precompute the MI clusters 
and query the \allSeeds seeds to obtain their cluster identifier.
Then, we use the WYB tag dataset to identify tagged clusters.
This step flags 58 seeds.
Of those, 57 are online wallets in exchanges.
The other is the only seed for the Razy family in the Ransomwhere dataset,
which really is an FBI address used to seize the Silk Road funds 
after its takedown.
We remove Razy from our datasets.

An intrinsic issue in tag databases is the limited coverage,
i.e., only a fraction of all MI clusters belonging to services will be tagged.
Indeed, when we perform an initial \emph{DD-OW+MI-DC} 
estimation for each cybercrime type,
the estimation for giveaway scams returns an
astronomical 6.1 Billion USD. This indicates that the WYB tag database likely missed some service clusters.
To identify untagged service clusters,
we run MI seed clusters with at least 1K addresses
through the exchange classifier provided by WYB.
The classifier identifies 8 of those 11 MI clusters as exchanges:
5 in the giveaway scams dataset, 
2 in the Ransomwhere dataset, and 
1 in the exchange scams dataset.
We consider the 9 seeds in these 8 clusters to be online wallets
for all other estimations.
Had we not identified these 9 online wallets,
we would have grossly overestimated the financial impact of
three cybercrime types:
\$6.1B for giveaway scams (503 times higher than our final estimation),
\$543M for ransomware (5 times higher), and
\$231M for exchange scams (226 times higher).

While only 0.7\% of seeds are online wallets, 
they can introduce huge overestimation. 
To avoid missing online wallets, we recommend complementing
tag databases with ML classifiers.

\paragraph{MI clustering impact.}
We use BlockSci to precompute the MI clusters 
and query each of the labeled seeds that are not online wallets 
to obtain their cluster.
The seeds belong to 968 clusters, 
with an average cluster size of 19.5 addresses, and 
a median size of 1 address.
Four clusters contain seeds from more than one group. 
One cluster holds the seeds for \emph{mekotio} and \emph{mekotion40}, 
which are known to be run by the same operators~\cite{wyb}.
Another holds Towerweb and Cryptohitman seeds and 
was already reported in~\cite{ransomwareClouston}. The other two clusters hold Jigsaw and Cryptowall, and 
TripleM and APT, respectively.
We have not found reports linking the groups in these two clusters. 
However, the small size of those two clusters 
(3 and 4 addresses, respectively) makes us think they may 
be true relations or incorrect labels on the seeds. 
Clustering FPs typically create a snowball effect that quickly 
grows the clusters, as we show next for the change address expansion.

\paragraph{Change address impact.}
We use BlockSci to precompute the MI+CA clusters 
and query each of the \groupSeeds labeled seeds from the 141 groups to 
obtain their cluster.
The \groupSeeds seeds belong to 251 clusters 
with an average cluster size of 2.2M addresses, 
and a median size of 2 addresses.
The higher average is due to
a huge cluster with 543M addresses. This cluster contains seeds from 91 of the 141 groups (64.5\%).
For 70 groups, all their seeds are in this cluster.
It also contains addresses tagged by WYB as belonging to a 
variety of different services such as 
exchanges, gambling sites, and mixers.
All other clusters with seeds contain one group at most,
except for one cluster that contains two 
malware families known to belong to the same operators
(\emph{mekotio} and \emph{mekotion40})~\cite{wyb}.

The fact that one cluster contains 91 unrelated groups 
and many unrelated services
indicates the change address heuristic is introducing FPs.
Because of this, we suggest that estimations do not use change address 
expansion. We further discuss this in Section~\ref{sec:discussion}.

\paragraph{Double-counting filter impact.}
To quantify the impact of the DC filter,
we perform two estimations on 
all \allSeeds seeds 
using the DD-OW+MI and DD-OW+MI-DC methodologies, respectively.
DD-OW+MI estimates a total revenue of 171,994.9777 BTC,
while DD-OW+MI-DC estimates 41,091.6373 BTC.
Thus, the DC filter reduces the BTC estimate by 76.1\%. 
This is much higher than the reduction observed in CryptoLocker,
as the impact of the DC filter quickly grows as the volume and 
fund movement increases.
In detail, the DC filter drops 43.7\% of deposits to the expanded set and 
only 80.8\% of the addresses in the expanded set
receive funds from outside the expanded set.

\subsection{Cybercrime Estimations}
\label{sec:cybercrimes}

All works in Table~\ref{tab:related} have estimated one or multiple 
groups of the same type of cybercrime. 
In this section, we instead compare revenue across cybercrimes. 
We perform all estimations 
at the same block height of 785,100 (April 12, 2023), 
using the same \emph{DD-OW+MI-DC} methodology, and 
the conversion rate on the day of each payment.
We select this methodology because we have shown that 
change address expansion introduces many false positives and 
we only have value and time filtering ranges for some 
ransomware families.

\begin{table}
\centering
\scriptsize
\begin{tabular}{l|r|r|r|r|r|r}
\hline
\textbf{Name} & \textbf{Seeds} & \textbf{OW} & \textbf{Clust} & \textbf{Addr.} & \textbf{BTC} & \textbf{USD} \\
\hline
R:Netwalker	& 66	& 0	& 65	& 329	& 3,131.7562	& \$28,262,120 \\ R:Conti	& 26	& 0	& 26	& 35	& 410.8025	& \$18,869,852 \\
R:REvil	& 7	& 0	& 6	& 77	& 369.5363	& \$12,540,689 \\ R:DarkSide	& 3	& 0	& 3	& 3	& 158.7086	& \$9,124,091 \\
R:Locky	& 7,036	& 0	& 2	& 7,094	& 15,396.6934	& \$7,833,613 \\
R:Ryuk	& 26	& 0	& 26	& 39	& 866.8069	& \$4,844,149 \\
R:RagnarLocker	& 1	& 0	& 1	& 1	& 414.0007	& \$4,547,241 \\
R:MountLocker	& 1	& 0	& 1	& 1	& 298.5000	& \$4,226,277 \\
R:BlackMatter	& 1	& 0	& 1	& 1	& 96.3863	& \$4,068,295 \\
R:Makop	& 1	& 0	& 1	& 996	& 316.0883	& \$3,853,713 \\
R:Egregor	& 9	& 0	& 9	& 9	& 197.9554	& \$3,129,599 \\
R:Tejodes	& 1	& 0	& 1	& 178	& 345.5527	& \$2,798,239 \\
P:Leancy	& 1	& 0	& 1	& 1	& 3,092.8482	& \$1,896,583 \\
R:CryptXXX	& 1	& 0	& 1	& 1,742	& 3,338.5464	& \$1,877,833 \\
R:DMALockerv3	& 9	& 0	& 3	& 177	& 1,833.9286	& \$1,757,464 \\
R:MedusaLocker	& 3	& 0	& 3	& 26	& 56.1546	& \$1,454,776 \\
R:HelloKitty	& 1	& 0	& 1	& 1	& 32.3529	& \$1,070,653 \\
R:Bitpaymer	& 1	& 0	& 1	& 1	& 90.0004	& \$1,058,347 \\ C:cliptomaner	& 1	& 1	& 1	& 1	& 38.7643	& \$1,004,397 \\
P:Cryptory	& 1	& 1	& 1	& 1	& 1,677.5167	& \$886,689 \\
\hline
Total	& 8,021	& 42	& 383	& 17,040	& 40,282.3312	& \$123,357,041 \\
\hline
\end{tabular}
\caption{Top-20 groups by USD financial impact on April 12, 2023 
using a \emph{DD-OW+MI-DC} estimation. 
Ransomware families are prefixed by R:, 
clipper families by C:, and
Ponzi schemes by P:.
The last row captures total revenue of all 141 labeled groups.
}
\label{tab:ransomwhere}
\vspace{-0.5cm}
\end{table}

\paragraph{Cybercrime comparison.}
The right side of Table~\ref{tab:cybercrime} summarizes the BTC and USD 
estimations for the 6 cybercrimes.
The largest estimated revenue is \$115.5M for ransomware,
followed by giveaway scams (\$12.2M),
Ponzi schemes (\$5.0M),
clippers (\$2.8M), 
sextortion scams (\$1.5M),
and exchange scams (\$1.0M).
Giveaway scams collect lower BTCs (386.0907) than 
Ponzi schemes (7,956.3913) and clippers (891.5331). 
However, their USD financial impact on victims is larger due to their 
seeds being collected in 2022, 
when the conversion rate BTC-USD was higher than for older seeds.

\paragraph{Group comparison.}
We also estimate the \numGroups groups. 
Table~\ref{tab:ransomwhere} shows the top-20 groups by USD revenue and 
the \numGroups groups are detailed in 
the Appendix of our extended version~\cite{gomez2023cybercrimeTR}. 
We prefix each name with R: for ransomware families, 
C: for clipper families, and 
P: for Ponzi schemes.
Of the top-20 groups, 17 are ransomware families,
two are Ponzi schemes and one is a clipper family.
The highest revenue is for the \emph{Netwalker} ransomware 
which collects \$28.2M.
The highest revenue among Ponzi schemes 
is \emph{Leancy} with \$1.8M and
the highest revenue among clippers is \emph{Cliptomaner} with \$1.0M.
The average revenue per operation is \$874K,
and the median is \$17.5K. 
There are 19 groups 
with revenues higher than one million USD and
14 groups (13 ransomware families and one Ponzi scheme) with revenue below
\$20, likely due to limited seed coverage for those groups.

Similar to what we observe in the cybercrime comparison, 
higher BTC revenue does not necessarily imply a higher financial
impact on victims, measured using fiat currency,
due to the wide oscillations of the Bitcoin conversion rate. 
For example, Locky collected the most BTCs (15,396.6934).
However, its financial impact measured in USD is below that of DarkSide, 
which received nearly two orders of magnitude fewer BTCs (158.7086).
The reason is that
Locky was active from the beginning of 2016 to mid-2017
when the value of 1 BTC was \$780 on average. In contrast, DarkSide was active during the first half of 2021,
when the conversion rate ranged from \$29k to \$61k (\$55.2k on average).
This example highlights the importance of using the conversion rate at 
the time of the payment for more accurate estimations
of the financial impact on the victims.

The top-4 earners in Table~\ref{tab:ransomwhere} 
(Netwalker~\cite{netwalker}, Conti~\cite{conti}, REvil~\cite{revil}, 
DarkSide~\cite{darksideDoS})
leverage the ransomware-as-a-service (RaaS) model,
where the ransomware operators recruit affiliates to take care of 
the infections~\cite{raas}. 
Affiliates are paid a fraction of the victim's ransom payment or a fixed fee.
Outsourcing infections to affiliates helps ransomware gangs target more victims.
These families focus on high-value targets such as large companies and 
hospitals, 
e.g., DarkSide was behind the Colonial Pipeline attack~\cite{darkside}.
They often use different ransom values for each victim, 
threaten victims to release their data publicly, and may even 
perform double extortion, 
i.e., demand a second payment for not leaking the data.
Targeted victims may not want to disclose the attacks, 
making it difficult to obtain seeds.
For example, there are reports of DarkSide having at least 90 victims in the 
US~\cite{darkside}, 
but we only have 3 seeds for that group and MI clustering does not reveal 
additional payment addresses.
We examine the impact of the lack of coverage in 
Section~\ref{sec:coverage}.

\subsection{Multi-Input Evasion}
\label{sec:evasion}

This section examines how useful MI clustering on the seeds is, 
and whether cybercriminals may be using evasive techniques to defeat it.
For this, we examine the withdrawal transactions from the seeds 
of the \numGroups groups, and their MI clusters.
For each group, we first remove seeds identified as online wallets.
Then, we compute
the total number of withdrawals from the remaining seeds and
the fraction of \emph{1-to-n} withdrawals.
Next, we compute the expanded set of each group 
by doing the union of the addresses in the MI clusters of the seeds 
that are not online wallets and
recompute the above numbers for all withdrawals from the expanded set 
of each group.

\begin{figure}[t]
\centering
\includegraphics[width=.40\textwidth]{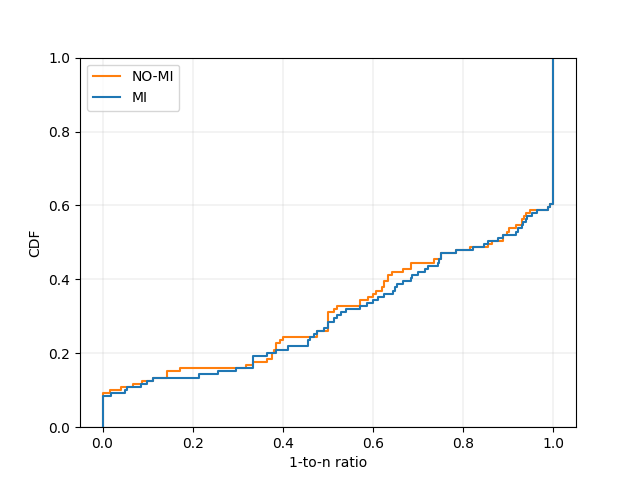}
\caption{Cumulative distribution of the proportion of 
withdrawals with one input address 
(1-to-n) over all the withdrawals
across the \numGroups groups for both withdrawals of seeds (orange line)
and withdrawals of their MI clusters (blue line).}
\label{fig:cdf}
\vspace{-0.4cm}
\end{figure}

Figure~\ref{fig:cdf} shows the cumulative distribution of 1-to-n withdrawals 
across the \numGroups groups.
40\% of the groups exclusively move funds from their payment addresses 
using MI-defeating 1-to-n withdrawals.
Three-quarters of the groups use at least 50\% of 1-to-n withdrawals, 
indicating their preference for such MI-defeating withdrawals.
Only 9\% of the groups do not use 1-to-n withdrawals at all, 
but most of these have not withdrawn any funds yet.
The fact that 40\% of the groups exclusively 
use MI-defeating withdrawals and 75\% use them more than half of the time, 
likely indicates that cybercriminals are aware of the limitations
of MI clustering and are actively taking steps to defeat it. 
We discuss how to address this evasion in Section~\ref{sec:discussion}.

\section{Coverage Impact}
\label{sec:coverage}

A recurring issue in cybercrime estimations is that 
they start from a small number of seeds (a median of one seed per group) 
and it is not 
clear how many other payment addresses with victim deposits may exist.
Expansions are used to try to discover missing payment addresses, 
but as shown in Section~\ref{sec:methodimpact} their effectiveness is limited.
As far as we know, no prior work has tried to quantify the impact of 
the lack of coverage in the estimations, due to the complexity 
of obtaining the needed vantage point.
In this section, we quantify this issue for the first time using 
two novel techniques that allow us to obtain very high coverage, 
possibly nearly complete, on the DeadBolt server ransomware~\cite{deadbolt}.
We introduce DeadBolt in Section~\ref{sec:deadbolt}, 
describe our DeadBolt datasets in Section~\ref{sec:datasets}, 
present our techniques to expand DeadBolt's coverage in 
Section~\ref{sec:covincrease}, and compare the estimations 
obtained from different vantage points in Section~\ref{sec:covcomparison}.

\subsection{DeadBolt}
\label{sec:deadbolt}

On January 25th, 2022, BleepingComputer first reported a 
new server ransomware strain that called itself \emph{DeadBolt} and 
was spreading by exploiting vulnerabilities
in network-attached storage (NAS) devices~\cite{deadbolt}. 
DeadBolt encrypted specific data directories and file extensions 
and hijacked the login page of the NAS to display a 
ransom note titled ``WARNING: Your files have been locked by DeadBolt''. 
The ransom note requested a Bitcoin payment of 0.03 BTC (1,100 USD at the time)
to be sent to a Bitcoin payment address to obtain the decryption key.
In mid-June 2022, DeadBolt introduced a new ransom value of 0.05 BTC, 
possibly due to a drop in the BTC-USD conversion rate at that time, 
and both amounts are used therefore.

\begin{figure}[t]
  \centering
  \includegraphics[width=.7\linewidth]{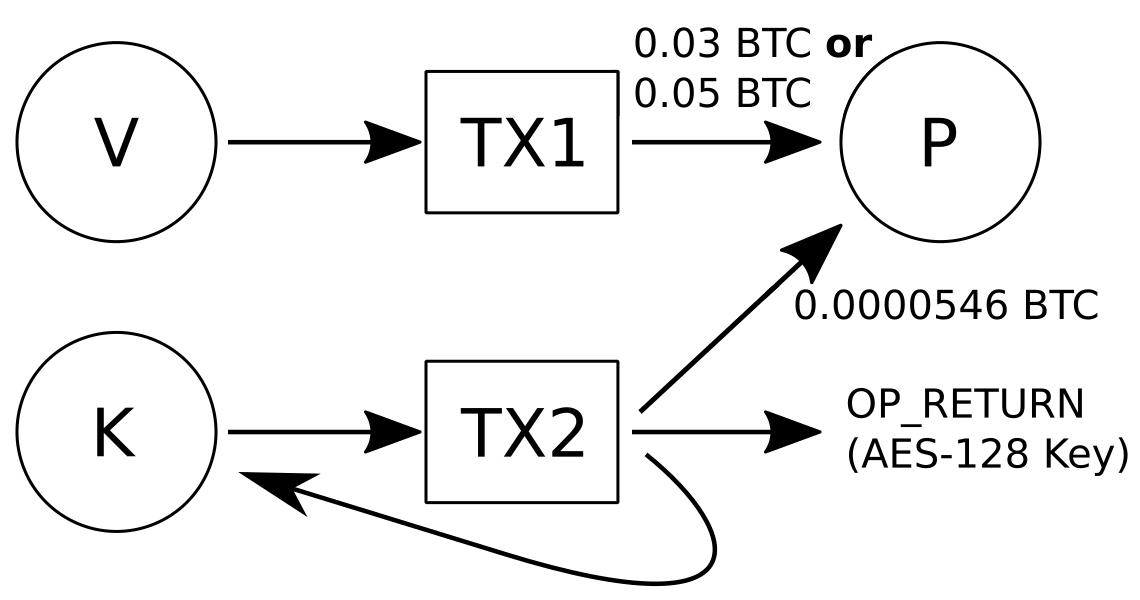}
  \caption{DeadBolt key release transactions.
First, the victim deposits the ransom of 0.03 BTC or 0.05 BTC to the payment 
address.
Then, the cybercriminals release the decryption key to the blockchain 
through a deposit to the payment address.}
  \label{fig:deadbolt}
\end{figure}

One novel feature in DeadBolt is that after the victim pays the ransom, 
the decryption key is automatically posted to the Bitcoin blockchain 
by the DeadBolt operators.
Figure~\ref{fig:deadbolt} illustrates the two transactions involved. 
First, the victim (address $V$) pays the ransom of 0.03 BTC or 0.05 BTC to the 
payment address $P$.
Then, the DeadBolt operators perform a deposit 
of value 0.0000546 BTC (\$1.2) from a \emph{key release address} $K$ 
to the payment address $P$. 
This second transaction has an OP\_RETURN output that stores the 
AES-128 decryption key on the blockchain. 

On October 2022, the Dutch National Police
tricked the DeadBolt operators into handing over 155 decryption keys 
by performing ransom payments with a very small fee
when the blockchain was heavily congested, 
and canceling the payment transactions before they appeared 
in a block~\cite{deadboltHack}.
Since the decryption key was automatically posted to the blockchain, 
without waiting for the payment transaction to appear in a block, 
the police recovered 155 keys before the attackers realized.
A webpage was set up that given one of the 155 addresses, 
outputs its decryption key~\cite{deadboltKeys}.
After that event, the DeadBolt operators maintained 
the key release procedure in Figure~\ref{fig:deadbolt} 
but adjusted their processing 
to wait for the payment transaction to appear in a block 
before releasing the decryption key.
From this moment on, the time difference between 
victim payments and key release transactions starts varying
significantly, reaching even 3 days in some cases,
indicating the key release transactions are now manually triggered.

\begin{table}[t]
	\small
	\centering
	\scalebox{0.9}{
		\begin{tabular}{l|r|r|r|r|r|r|r|r}
			\hline
			\textbf{Engine} & \textbf{Events} & \textbf{IP} & \textbf{Port} & \textbf{ASN} & \textbf{CC} & \textbf{Notes} & \textbf{Addr.} & \textbf{Seeds} \\
			\hline
			\textbf{Shodan} & 23,413          & 9,999       & 171           & 769          & 89          & 4,938          & 4,940          & 64               \\
			\textbf{Censys} & 15,886          & 6,147       & 66            & 671          & 87          & -              & 4,014          & 49              \\
			\hline
			\textbf{All}    & 39,299          & 11,282      & 199           & 780          & 98          & 4,938          & 4,997          & 64              \\
			\hline
		\end{tabular}
	}
	\caption{DeadBolt datasets summary.}
	\label{tab:summary_deadbolt}
	\vspace{-0.4cm}
\end{table}

\ignore{ 	\begin{table}[t]
		\small
		\centering
		\scalebox{0.9}{
			\begin{tabular}{lrrrrrrrrrrr}
				\hline
				\textbf{Engine} & \textbf{Events} & \textbf{IP} & \textbf{Port} & \textbf{ASN} & \textbf{CC} & \textbf{Notes} & \textbf{Addr.} & \textbf{Seeds} \\
				\hline
				\textbf{Shodan} & 4,608           & 3,525       & 30            & 553          & 81          & 2,834          & 2,836          &                \\
				\textbf{Censys} & 4,518           & 2,891       & 46            & 515          & 77          & -              & 2,674          &                \\
				\hline
				\textbf{All}    & 9,126           & 4,203       & 52            & 585          & 88          & 2,834          & 3,135          &                \\
				\hline
			\end{tabular}
		}
		\caption{DeadBolt datasets summary.}
		\label{tab:summary_deadbolt}
	\end{table}
}

\ignore{ 	\begin{table}[t]
		\small
		\centering
		\scalebox{0.9}{
			\begin{tabular}{lrrrrrrrrrrr}
				\hline
				\textbf{Engine} & \textbf{Events} & \textbf{IP} & \textbf{Port} & \textbf{ASN} & \textbf{CC} & \textbf{Notes} & \textbf{Addr.} & \textbf{Seeds} \\
				\hline
				\textbf{Shodan} & 7,714           & 5,756       & 104           & 715          & 88          & 4,497          & 4,499          & 50             \\
				\textbf{Censys} & 6,475           & 4,159       & 56            & 642          & 85          & -              & 3,844          & 40             \\
								\hline
				\textbf{All}    & 14,189          & 6,516       & 134           & 747          & 98          & 4,497          & 4,769          & 54             \\
				\hline
			\end{tabular}
		}
		\caption{DeadBolt datasets summary.}
		\label{tab:summary_deadbolt}
	\end{table}
}

\subsection{DeadBolt Datasets}
\label{sec:datasets}

DeadBolt exploited network-facing vulnerabilities in the NAS 
and hijacked the NAS login page to display the ransom note.
A NAS had to be connected to the Internet to get infected.
Thus, the ransom note was (potentially) visible to Internet scanners.
However, Internet scanners may not observe all victims 
since an infected NAS may have been disconnected or cleaned 
before being scanned.  
We obtain data about DeadBolt infections from two Internet scanners:
Censys~\cite{censys}, and Shodan~\cite{shodan}.
The datasets are summarized in Table~\ref{tab:summary_deadbolt}.

\paragraph{Censys.}
Censys publishes servers infected with DeadBolt in a
Google Datastudio~\cite{deadboltCensysDatastudio}.
Each entry in the dataset represents an infection and contains
the IP address,
the port number,
the country code (CC) and ASN of the IP address,
the payment address,
the ransom amount, and
the DeadBolt variant.
The raw content of the ransom notes is not available.
We collect Censys data for four months,
from \collectionStart, until \collectionEnd.
We obtain 4,014 DeadBolt payment addresses from 6,147 IP addresses.
The large number of payment addresses likely indicates each infected
server is given a different one.
The number of IP addresses is larger than the number of
payment addresses likely due to infected servers that change
IP addresses.

\paragraph{Shodan.}
We identify DeadBolt events in Shodan by querying its API
with the dork
\texttt{http.title:"Your files have been locked by DEADBOLT."}.
Each query returns events from the last 30 days,
so we query once a month, from \collectionStart, until \collectionEnd.
From each event, we extract
the timestamp,
IP address,
port number,
country code and ASN of the IP address, and
the ransomware note in HTML format.
We obtain 4,938 distinct ransom notes from 9,999 IP addresses.
We extract 4,940 payment addresses from the ransom notes using
the \emph{iocsearcher} tool~\cite{iocsearcher}.

\begin{figure}[t]
	\begin{minipage}[t]{0.49\linewidth}
		\centering
		\includegraphics[width=\textwidth]{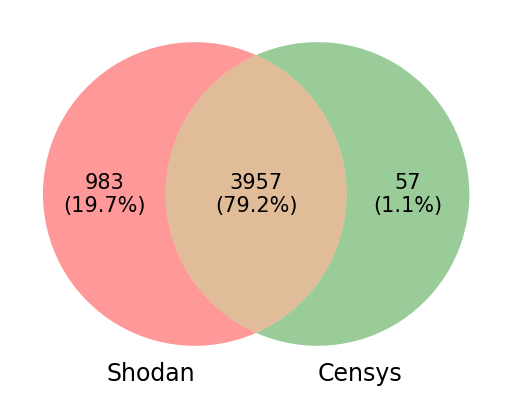}
		\subcaption{Bitcoin addresses.}
    \label{fig:venn_diagram_btc}
			\end{minipage}
	\hfill
	\begin{minipage}[t]{0.49\linewidth}
		\centering
		\includegraphics[width=\textwidth]{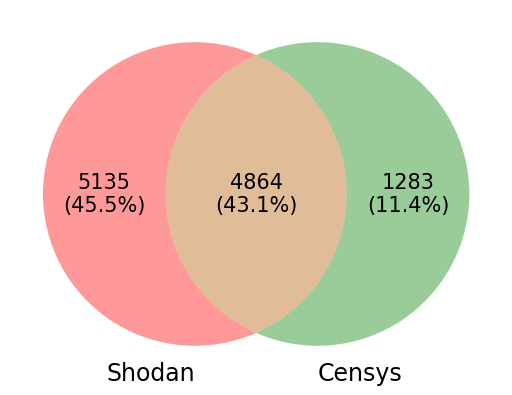}
		\subcaption{IP addresses.}
    \label{fig:venn_diagram_ip}
		\label{fig:2}
	\end{minipage}
	\caption{Overlap in DeadBolt datasets.}
	\label{fig:venn_diagrams}
	\vspace{-0.4cm}
\end{figure}

Both scanners observe different sets of infected servers,
largely due to their scanning happening at different times.
Figure~\ref{fig:venn_diagram_btc} shows that
79.2\% of the payment addresses appear in both datasets,
19.7\% only in Shodan, and
1.1\% only in Censys.
Figure~\ref{fig:venn_diagram_ip} shows that
43.1\% of the IP addresses appear in both datasets,
45.5\% only in Shodan, and
11.4\% only in Censys.
Thus, our Shodan dataset has slightly better coverage on DeadBolt,
but the set of payment addresses and infected IPs observed from
both vantage points significantly overlaps.
The larger overlap in payment addresses is likely due to
servers that appear multiple times on different IP addresses.

\subsection{Increasing the Coverage}
\label{sec:covincrease}

Of the 4,997 DeadBolt payment addresses collected from Shodan and Censys,
only 64 addresses
had received deposits as of April 12, 2023.
For 30 of those 64 seeds, the Bitcoin ledger does not show the victim's
payment, but only the key release transaction,
all of them on October 13, 2022.
These 30 addresses are part of the 155 addresses for which
the Dutch police recovered their decryption key.
Since the ransom payment was withdrawn after the key was released,
the payment transactions do not appear on the ledger.
The fact that our datasets only have 30 of the 155 addresses for which
the Dutch police recovered the keys indicates that we only have
partial coverage. 
None of the seeds have withdrawals.
Thus, the received payments have still not been moved and
expansions can not reveal new addresses.
In the remainder of this section, we present two novel techniques
that leverage unique characteristics of DeadBolt,
which allow us to obtain very high coverage, possibly nearly complete, 
on the payments performed by DeadBolt victims.

\paragraph{Key release analysis.}
All key release transactions to the 64 seeds have the same format, 
illustrated in Figure~\ref{fig:deadbolt},
with one input key release address 
and three output slots that
correspond to
(a) the payment address that receives 0.0000546 BTC, 
(b) the input key release address which is used as change address and
(c) the OP RETURN script that stores the decryption key on the blockchain.

Critically, all key release transactions for the 64 seeds
originate from one of two key release addresses:
\addr{bc1qh6}\footnote{
bc1qh6pku7gg2d6pw87z3t4f6d4rk6c48ajvsmfjjl
}
or \addr{bc1q62}\footnote{
bc1q62rjm9a82s3qmjzffc6uyytw25p3fppftl5zpd
}.
The reuse of key release addresses to post the decryption keys of many victims
allows us to identify ransom payments to previously unknown 
payment addresses (i.e., not in our datasets) that correspond to 
infected servers the Internet scanners did not observe
(e.g., NAS devices disconnected from the Internet right after infection).
For this, we retrieve all withdrawals
from the two key release addresses.
All these transactions match the expected pattern,
so their output addresses receiving 0.0000546 BTC are
(potentially unknown) payment addresses.
This technique identifies 2,481 payment addresses,
of which 2,418 (97.5\%) are previously unknown.

The payment addresses with deposits we know 
at this point contain 154 addresses with
only the key release transaction, but no victim payment.
This almost matches the 155 keys the Dutch Police were able to extract
in October 2022.
Still, there is one missing address indicating we may still be missing
some victim payments.
One possible reason for this is that there could exist other key release
addresses that we have not observed in our datasets.

\paragraph{Key release signature.}
We propose a novel technique to identify additional DeadBolt key release
addresses.
We leverage that DeadBolt key release transactions
are quite distinct, build a signature for them, and
scan all transactions in the Bitcoin ledger for over 15 months to search
for unknown key release addresses.
Given a transaction, our signature checks
that it has one input and three outputs slots,
where the output slots correspond (in any order) to
(1) an OP RETURN address,
(2) an address receiving exactly 5460 satoshis (0.0000546 BTC), and
(3) an address that is the same as the sending address.
We apply the signature to all Bitcoin transactions 
from block 716,591 (2022-01-01)
up to block 785,100 (2023-04-12).
The search takes 16 minutes to complete.
We decode the OP RETURN payload data from the matched transactions and
observe that several decoded values start with the string \textit{omni},
indicating that they are part of the OMNI protocol~\cite{omni}.
After filtering the OMNI transactions,
all remaining transactions originate from 3 addresses,
two of them are the known key release addresses, and
the third one (\addr{bc1q3g}\footnote{bc1q3guvg2yp5mzmf7hnfr7zlg2unah9t6mjwyky72})
is a previously unknown key release address.
The newly discovered key release address has delivered the decryption
key to 18 (previously unknown) payment addresses.
Of those, 17 receive the expected payments of 0.03 BTC or 0.05 BTC.
The other address only received the key release transaction,
but no victim payment.
This address receives the missing key of those recovered by the Dutch police,
as confirmed by using the public Web service.

The fact that we have scanned all Bitcoin transactions since DeadBolt
started to operate and have not identified further key release addresses,
along with the fact that we observe all 155 keys recovered by
the Dutch Police, give us strong confidence that we have achieved
nearly perfect coverage on DeadBolt's victim payments.
Of course, our key release signature assumes a unique
key release transaction pattern,
so we could miss victim payments if multiple patterns exist.

\ignore{
	Indeed, the lack of a key release transaction
	may avoid the detection of a valid payment address
	by our key release detection method.
	Apart from this address found by MI clustering,
	another two addresses were missed: \addr{bc1qac} and \addr{bc1qgj}.
			However, both addresses were part of the Shodan-Censys dataset.
	If we consider these three addresses as false negatives,
	the false detection rate (FDR) of our approach would still be very low (0.12\%).
}

\begin{table}
\centering
\resizebox{\columnwidth}{!}{\begin{tabular}{l|r|r|r|r|r|r}
\hline
	\textbf{Collection} & \textbf{Coverage} & \textbf{Seeds} & \textbf{Addr.} & \textbf{Dep.} & \textbf{BTC} & \textbf{USD} \\
\hline
	Censys			&   2.0\%	&    50	&    50	&   86	&  2.45686936	&    53,151.89	\\
	Shodan			&   2.6\%	&    66	&    66	&   114	&  2.82608808	&    63,077.10	\\
	Censys+Shodan		&   2.6\%	&    66	&    66	&   114	&  2.82608808	&    63,077.10	\\
\hline
	Key rel. analysis	&  99.2\%	& 2,484	& 2,485	& 2,596	& 97.83718238	& 2,453,532.39	\\
	\hline
	Key rel. signature	& 100.0\%	& 2,503	& 2,504	& 2,615	& 98.35008368	& 2,472,845.02	\\
	\hline
\end{tabular}
}
\caption{DeadBolt estimations from different vantage points.}
\label{tab:coverage}
\end{table}

\subsection{Coverage Comparison}
\label{sec:covcomparison}

Table~\ref{tab:coverage} summarizes the coverage and
revenue estimation on DeadBolt
from different vantage points:
using only Censys, only Shodan, both Censys and Shodan, and
extending our coverage using the key release analysis and
the key release signature cumulatively.
The table shows the fraction of DeadBolt addresses, 
i.e., payment addresses plus key release addresses,
found at each step (Coverage),
the number of addresses with deposits (Seeds),
the number of addresses found after applying MI clustering (Addr.),
the number of deposits to the addresses (Dep.),
the BTC estimation on block height 785,100 using a \emph{DD-OW+MI-DC}
methodology, and the USD estimation using the conversion rate at the time of
each payment.

\begin{figure}[t]
  \centering
  \includegraphics[scale=.5]{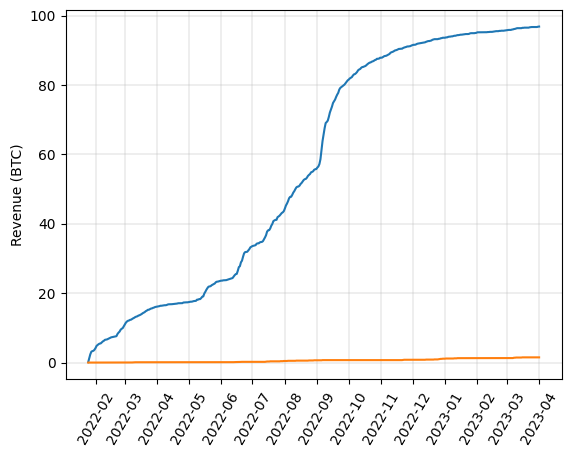}
  \caption{Cumulative revenue for DeadBolt over time using
full coverage (blue line) and only scan engines (orange line).}
  \label{fig:deadbolt_profit_acc_day}
\end{figure}

\begin{figure}[t]
  \centering
  \includegraphics[scale=.5]{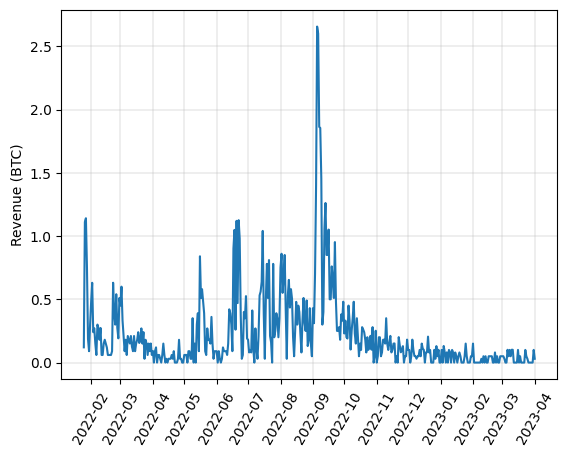}
  \caption{DeadBolt revenue per day.}
  \label{fig:deadbolt_profit_day}
\end{figure}

\ignore{
  \begin{figure}[t]
    \centering
    \includegraphics[scale=.5]{fig/deadbolt_profit_week.png}
    \caption{DeadBolt payments per weeks}
    \label{fig:deadbolt_profit_week}
  \end{figure}

  \begin{figure}[t]
    \centering
    \includegraphics[scale=.5]{fig/deadbolt_profit_month.png}
    \caption{DeadBolt payments per months}
    \label{fig:deadbolt_profit_month}
  \end{figure}

  \begin{figure}[t]
    \centering
    \includegraphics[scale=.5]{fig/deadbolt_profit_acc_week.png}
    \caption{DeadBolt accumulated payments per weeks}
    \label{fig:deadbolt_profit_acc_week}
  \end{figure}

  \begin{figure}[t]
    \centering
    \includegraphics[scale=.5]{fig/deadbolt_profit_acc_month.png}
    \caption{DeadBolt accumulated payments per months}
    \label{fig:deadbolt_profit_acc_month}
  \end{figure}
}

Had we focused only on the payment addresses observed in the union of
the Censys and Shodan datasets,
we would have only identified 66 addresses 
(64 payment addresses and 2 key release addresses),
2.6\% of all DeadBolt addresses finally identified.
Of those addresses, only 34 had victim payments and
the MI expansion fails to identify additional addresses
since the funds have not yet been moved from those 34.
Thus, we would have estimated a very modest revenue of 2.826 BTCs or
\$63,077.

Instead, by applying our two coverage-expanding techniques,
we identify 38 times more seeds.
The estimation jumps to 98.350 BTC,
35 times higher than the 2.826 BTC.
The estimated USD revenue is \$2.47M, 
39 times higher than the \$63,077 using the Internet scanners datasets.
Figure~\ref{fig:deadbolt_profit_acc_day} shows the difference in cumulative 
DeadBolt revenue using both vantage points and
Figure~\ref{fig:deadbolt_profit_day} the daily BTC amounts
received by DeadBolt using our most complete coverage.

In summary, the estimation using only the Shodan and Censys datasets
would have been 2.5\% of the total USD estimation,
highlighting the huge impact of the (lack of) coverage in the estimation.
This in turn means our estimations in Section~\ref{sec:cybercrimes}
may significantly underestimate the financial impact of Bitcoin-related
cybercrimes.

\section{How To Estimate}
\label{sec:howto}

It is widely accepted that estimated cybercrime Bitcoin
revenue is a lower bound on the actual revenue due to the impact of the
lack of coverage on the seeds.
Indeed, we have quantified the coverage impact to be very significant
on the DeadBolt ransomware.
However, our results show that some methodology steps,
e.g., handling online wallets in services
and using change address heuristics,
may introduce huge overestimation.
In some cases, such methodological errors
can outweigh the underestimation caused by the lack of coverage,
thus overshooting the actual revenue.

This section presents our recommendations on how to perform estimations.
At a high level, we recommend performing two BTC revenue estimations:
DD-DC and DD-OW+MI-DC. 
First, as a conservative estimation,
we recommend authors provide the sum of direct deposit on seeds 
excluding deposits where seeds appear both in 
the input and output slots (DD-DC).
This estimation does not use expansions. 
Thus, it is easy to compute without specialized platforms,
e.g., deposit transactions to seeds are available from
blockchain explorer webpages and APIs,
and the double counting filter only requires examining if the seeds 
appear in the input slots of the deposits.
On the other hand, DD-DC may significantly underestimate revenue if 
seed coverage is low.
Overestimation only happens if seeds are incorrectly labeled or 
reused for other purposes.
We discuss the latter below in the discussion of value and time filtering.

As a tighter estimation, we recommend using DD-OW+MI-DC,
which we used for the comparison in Section~\ref{sec:cybercrimes}.
This estimation uses the MI clustering expansion
(with special handling for seeds that are online wallets in services), 
avoids change address expansion, and
removes duplicate deposits. 
We have shown that identifying online wallets in services like exchanges 
is critical to prevent huge overestimation when using MI clustering.
For such identification, it is possible to use
publicly available tag databases like those
in WYB~\cite{wyb} and GraphSense~\cite{graphsense}.
However, tag datasets have limited coverage. 
For example, our initial giveaway scams estimation was an astronomical 
6.1 Billion USD (503 times higher than the final estimation) 
due to seeds in five exchanges not identified by the WYB tags.
Because of that, an ML classifier such as the one in WYB
should complement tag databases.
An alternative is using heuristics to exclude outlier clusters. 
However, it is hard to select reliable thresholds
on cluster size (e.g., Locky's MI cluster has 7,094 addresses) or 
BTC volume (given the highly volatile BTC conversion rate).
The best option may be to exclude clusters with multi-million USD revenue, 
but this may prevent the analysis of large operations.
To prevent overestimation, 
MI clustering should use detection heuristics for CoinJoin mixing 
transactions~\cite{blocksci,cookieGoldfeder},
available in open-source platforms~\cite{blocksci,graphsense}.
This estimate avoids the change address expansion due to its 
potentially high false positives,
which prevented us from obtaining some results in 
Section~\ref{sec:cryptolocker}. 
Recently, Kappos et al.~\cite{kappos22peel} proposed a new change address 
heuristic with lower false positives than prior heuristics.
This heuristic has not been used for estimations yet,
but we would like to evaluate it in future work.
This estimate also avoids double counting deposits, 
which we have shown can introduce overestimation.
In particular, we have measured that applying the double-counting
filter on all \allSeeds seeds
removes 76\% of the estimated revenue.

\paragraph{Value and time filtering.}
The proposed estimations do not use value and time filtering 
for several reasons. 
First, these filters are helpful only if payment addresses are reused
for multiple purposes (e.g., different campaigns or cybercrimes) 
and the estimation targets just one
(e.g., a specific campaign). 
Second, they are group-specific, and the selection of a proper
filtering range is affected by the group's coverage.
In addition, value filtering does not apply to cases where the 
payment amount is not pre-determined
(e.g., clippers and giveaway scams) and 
when each victim is assigned a different amount
(e.g., some ransomware families).
If authors believe value and time filtering 
(or any other group-specific filtering)
should be applied, 
we recommend providing estimations with and without those filters.

\paragraph{Other blockchains.}
Three surveyed works also collect seeds for 
other blockchains such as Cardano, Ethereum, Monero, and 
Ripple~\cite{cybercriminalLee,xia2020characterizing,li2023giveaway}.
We have focused on estimating Bitcoin revenue for two reasons.
First, prior works exclusively apply a DD estimation on other blockchains.
Thus, there are no methodologies to compare. 
Besides, \tool currently only supports Bitcoin.
To estimate the revenue of a cybercriminal activity 
that solicits victim payments on multiple blockchains, 
we recommend performing a separate DD-DC estimation
on each blockchain. 
For each blockchain, deposits to the seeds can be obtained 
using the webpage or API of blockchain explorers and 
the double counting filter only requires checking if seeds appear 
among the inputs of each deposit.
We expect future research to investigate more sophisticated 
estimation methodologies for other blockchains (e.g., novel expansions).

\paragraph{Conversion rate.}
The above recommendations focus on estimating BTC revenue. 
To estimate the financial impact on victims using fiat currency 
(e.g., USD) we recommend applying the BTC to fiat conversion rate
on the day each payment was received.
This conversion accounts for a victim's loss at the time it happened. 
Conversions using a later fixed date are difficult to interpret
due to the high variability of the BTC conversion rate over time.
In particular, we have shown how older cybercriminal campaigns that collected 
high amounts of BTC (e.g., Locky) 
had a lower USD financial impact on victims 
than more recent campaigns that collected fewer BTCs (e.g., DarkSide).

\paragraph{Enabling replicability.}
Bitcoin transactions are immutable and public,
so in theory, estimation results should be easy to replicate.
In practice, it is often not so due to missing information.
To make results replicable, we recommend that estimation works 
estate their estimation target 
(e.g., a cybercrime type, a cybercriminal group, or a specific campaign) and
release their seeds and the block height used for the estimation.
If applicable, they should also release filtering ranges and 
clustering results. 
We also recommend releasing other payment addresses that have not
received deposits yet, as that could change at higher block heights.
If using expansions, authors should mention which platform they are using
to implement them and release their code if custom.
If using change address expansion,
authors should specify the specific variant employed.

\section{Further Discussion}
\label{sec:discussion}

This section discusses other insights and future improvements.

\paragraph{Evasion.}
MI clustering fails to find new addresses for 40\% of the \numGroups groups, 
arguably indicating cybercriminals are actively evading this expansion. 
Huang et al.~\cite{trackingHuang} addressed this issue 
for the Cerber ransomware 
by observing that the seeds sent the funds to aggregators, 
on which MI clustering could be successfully applied.
We believe their approach can be generalized using recent platforms 
for tracing cybercriminal flows~\cite{wyb}. 
In particular, it may be possible to perform a 1-step forward exploration 
from the seeds and apply heuristics to determine if the discovered addresses 
belong to the cybercriminals. 
For example, if the attackers use 1-to-1 withdrawals then the 
no change heuristic~\cite{nochange} would determine 
the destination address belongs to the seed owners.
Cybercriminals can also evade the estimation by hiding their payment addresses.
For example, some ransomware families do not provide 
the payment address and value in their ransom notes, 
but only a contact email or IM address~\cite{esxiargs}. 
That extra level of indirection complicates the collection of seeds.

Another common evasion is the use of mixers. 
While their goal is to obfuscate the destination of funds, 
mixing transactions can also introduce estimation errors.
First, undetected mixing transactions may lead to overestimation
with MI clustering by bringing into the expanded set unrelated addresses
from other entities.
Second, undetected mixing transactions that deposit to the expanded set 
may originate from the same entity that owns the expanded set addresses, 
but will not be removed by the DC filter, thus introducing overestimation.
We use mixer tags available in WYB~\cite{wyb}
and CoinJoin detection heuristics
implemented by BlockSci~\cite{blocksci,cookieGoldfeder}
to address both issues.
However, we acknowledge that both approaches are incomplete.
Heuristics in BlockSci target CoinJoin, but there exists
a variety of other mixing protocols (e.g., CoinShuffle~\cite{coinshuffle}) and 
mixing techniques~\cite{towardswu}.
Tags can only identify known mixers and may miss  
mixing transactions in those, 
e.g., if mixing commission fees are not aggregated.
Future work should develop new mixing detection techniques that can 
be added to blockchain analysis platforms.

\paragraph{Generality of DeadBolt results.}
Our two coverage-increasing techniques are specific to DeadBolt.
Thus, the coverage impact may differ for other groups.
We believe the coverage impact will likely be larger in other cases
because Internet scanners should have better coverage on server ransomware
than most security vendors would on a malware family.
Furthermore, our two techniques
generalize beyond the DeadBolt coverage measurement.
The key release analysis technique is an instance of doing 
one backward exploration step
from a payment address to identify a key release address,
followed by one forward exploration step from the key release address
to discover other payment addresses.
It is related to the exploration Huang et al.~\cite{trackingHuang} performed 
for Cerber, although that case was one forward step from the seeds 
followed by one backward step.
We believe such forward+backward or backward+forward exploration
will be critical to improving estimation methodologies
(e.g., to address MI clustering evasion).
The key release signature technique builds a signature for a 
distinctive malicious transaction and scans all blockchain transactions 
in a time range to identify more transactions matching the signature.
We believe it should be possible to build signatures for 
other types of malicious Bitcoin transactions, such as
those used in C\&C signaling~\cite{wyb}.

\paragraph{DeadBolt conversion rate.}
Measuring the conversion rate of a cybercrime, 
i.e., the fraction of targets that pay,
is challenging and may require developing novel 
methodologies~\cite{spamalytics}.
The fact that DeadBolt's ransom notes are publicly available on the Internet 
allows us to identify infected servers (i.e., victims), 
obtain their payment addresses, and check what fraction received a 
deposit in the expected ranges
([0.029,0.031] or [0.049,0.051]).
Since we collect 4,997 payment addresses from Censys and Shodan,
and only 34 of those have received valid payments,
this indicates that 0.7\% of the victims paid the ransom.
This estimate relies on each server being assigned a unique payment address.
Among the 2,500 payment addresses with victim payments,  
only 13 (0.52\%) receive more than one deposit in 
the expected ranges. 
Thus, while not every server may have been assigned a unique address
(it is unlikely that the same victim payed multiple times), 
most seem to have their own, and the 0.7\% ratio is 
likely a good estimate of DeadBolt's conversion rate.
Given the nature of DeadBolt targets, 
i.e., NAS devices possibly storing much data (including data backups), 
this conversion rate may be very specific to DeadBolt and may not represent 
other (e.g., desktop) ransomware, which may have lower conversion rates.

\section{Conclusion}
\label{sec:conclusion}

We have presented the first systematic analysis
on the estimation of cybercrime bitcoin revenue. 
Our analysis has quantified the impact in the estimation of 
the methodology used and the limited seed coverage.
We have built a tool able to replicate the different methodologies and
have collected a dataset of \allAddr cybercrime payment addresses.
We have used them to quantify the impact of different methodology steps and
to compare the revenue obtained by 6 cybercrimes and 141 cybercriminal groups.
We show that some methodologies may produce large overestimation 
by introducing addresses unrelated to the campaign,
through undetected online wallets in services like exchanges,
or by double-counting deposits.
We have measured that for 40\% of the groups MI clustering fails to 
discover additional addresses,
and that ransomware dominates the Bitcoin-payment cybercrime scene
with a revenue almost 10 times larger than other cybercrimes.
For the first time, we have quantified the impact
of the lack of coverage in the estimation.
We propose two techniques to achieve possibly complete coverage of 
victim payments to the DeadBolt server ransomware. 
From our privileged vantage point, we estimate a revenue of \$2.47M,
39 times higher than estimated from the vantage point provided by two 
popular Internet scan engines.

\section*{Acknowledgments}
This work was partially funded by 
the Spanish Government MCIN/ AEI/10.13039/501100011033/
through grants 
TED2021-132464B-I00 (PRODIGY),
PID2022-142290OB-I00 (ESPADA), and 
PRE2019-088472. 
Those projects are co-funded by
European Union ESF, EIE, and NextGeneration funds.

{
	%\renewcommand{\baselinestretch}{1.0}
	%\footnotesize
	\bibliographystyle{abbrv}
	%bibliographystyle{formatting/splncs}
	%\bibliographystyle{plain --short=author --short=state --no-field editor --no-field publisher --no-field pages}
	%\vspace{-0.05in}

	\balance

	\bibliography{bibliography/paper}

\begin{thebibliography}{10}

\bibitem{ponziBartolettiDataset}
Bitcoinponzitool, 2018.
\newblock \url{https://github.com/bitcoinponzi/BitcoinPonziTool/}.

\bibitem{economicContiDataset}
On the economic significance of ransomware campaigns: A bitcoin transactions
  perspective (dataset), 2018.
\newblock
  \url{https://spritz.math.unipd.it/datasets/btcransomware/knowledge_base.zip}.

\bibitem{ransomwareCloustonDataset}
Ransomware in the bitcoin ecosystem | dataset extraction, 2019.
\newblock \url{https://github.com/behas/ransomware-dataset}.

\bibitem{spamsCloustonDataset}
Spams meet cryptocurrencies dataset, 2019.
\newblock \url{https://github.com/MatteoRomiti/Sextortion_Spam_Bitcoin}.

\bibitem{xia2020characterizingDataset}
Cryptocurrency exchange scams dataset, 2020.
\newblock \url{https://cryptoexchangescam.github.io/ScamDataset/}.

\bibitem{bitcoinabuse}
Bitcoin abuse, 2022.
\newblock \url{https://www.bitcoinabuse.com}.

\bibitem{wybrepo}
\blockchaintool, 2022.
\newblock \url{https://github.com/cybersec-code/watchyourback}.

\bibitem{nochange}
{Bitcoin Wiki: no change heuristic}, 2023.
\newblock
  \url{https://en.bitcoin.it/wiki/Privacy#Exact_payment_amounts_.28no_change.29}.

\bibitem{chainalysis}
Chainalysis, 2023.
\newblock \url{https://www.chainalysis.com/}.

\bibitem{coindesk}
Coindesk: Bitcoin, ethereum, crypto news and price data, 2023.
\newblock \url{https://api.coindesk.com/v1/bpi/historical/close.json}.

\bibitem{li2023giveawayDataset}
Double and nothing dataset, 2023.
\newblock \url{https://double-and-nothing.github.io/}.

\bibitem{iocsearcher}
iocsearcher, 2023.
\newblock \url{https://github.com/malicialab/iocsearcher}.

\bibitem{walletexplorer}
Wallet explorer, 2023.
\newblock \url{https://www.walletexplorer.com/info}.

\bibitem{deadbolt}
L.~Abrams.
\newblock {New DeadBolt ransomware targets QNAP devices, asks 50 BTC for master
  key}, January 2022.
\newblock \url{
  https://www.bleepingcomputer.com/news/security/new-deadbolt-ransomware-targets-qnap-devices-asks-50-btc-for-master-key/
  }.

\bibitem{evaluatingAndroulaki}
E.~Androulaki, G.~O. Karame, M.~Roeschlin, T.~Scherer, and S.~Capkun.
\newblock {Evaluating User Privacy in Bitcoin}.
\newblock In {\em Financial Cryptography and Data Security}, 2013.

\bibitem{bartoletti2021cryptocurrency}
M.~Bartoletti, S.~Lande, A.~Loddo, L.~Pompianu, and S.~Serusi.
\newblock {Cryptocurrency Scams: Analysis and Perspectives}.
\newblock {\em IEEE Access}, 9:148353--148373, 2021.

\bibitem{ponziBartoletti}
M.~Bartoletti, B.~Pes, and S.~Serusi.
\newblock {Data Mining for Detecting Bitcoin Ponzi Schemes}.
\newblock In {\em Crypto Valley Conference on Blockchain Technology}, June
  2018.

\bibitem{bijmans2019just}
H.~L. Bijmans, T.~M. Booij, and C.~Doerr.
\newblock {Just the Tip of the Iceberg: Internet-Scale Exploitation of Routers
  for Cryptojacking}.
\newblock In {\em ACM SIGSAC Conference on Computer and Communications
  Security}, 2019.

\bibitem{bitiodine}
Bitiodine, 2023.
\newblock \url{https://github.com/mikispag/bitiodine}.

\bibitem{ransomwhere}
J.~Cable.
\newblock {Ransomwhere: A Crowdsourced Ransomware Payment Dataset}, May 2022.

\bibitem{censys}
{Censys}, 2022.
\newblock \url{https://censys.io/}.

\bibitem{deadboltCensysDatastudio}
{Censys DeadBolt}, 2022.
\newblock \url{
  https://datastudio.google.com/reporting/f8d38b6c-9997-4bba-be93-19cf57d7371a/page/DcGtC
  }.

\bibitem{travelingChristin}
N.~Christin.
\newblock {Traveling the Silk Road: A measurement analysis of a large anonymous
  online marketplace}.
\newblock In {\em The World Wide Web Conference}, 2013.

\bibitem{economicConti}
M.~Conti, A.~Gangwal, and S.~Ruj.
\newblock On the economic significance of ransomware campaigns: {A} bitcoin
  transactions perspective.
\newblock {\em Computers \& Security}, 79:162--189, 2018.

\bibitem{netwalker}
N.~Coppinger.
\newblock {Netwalker Ransomware Guide: Everything You Need to Know}, February
  2022.
\newblock \url{https://www.varonis.com/blog/netwalker-ransomware}.

\bibitem{esxiargs}
M.~Ellzey and E.~Austin.
\newblock {The Evolution of ESXiArgs Ransomware}, February 2023.
\newblock \url{https://censys.io/the-evolution-of-esxiargs-ransomware}.

\bibitem{automaticErmilov}
D.~Ermilov, M.~Panov, and Y.~Yanovich.
\newblock {Automatic Bitcoin Address Clustering}.
\newblock In {\em IEEE International Conference on Machine Learning and
  Applications}, 2017.

\bibitem{deadboltHack}
S.~Gatlan.
\newblock {Police tricks DeadBolt ransomware out of 155 decryption keys},
  October 2022.
\newblock \url{
  https://www.bleepingcomputer.com/news/security/police-tricks-deadbolt-ransomware-out-of-155-decryption-keys/
  }.

\bibitem{cookieGoldfeder}
S.~Goldfeder, H.~A. Kalodner, D.~Reisman, and A.~Narayanan.
\newblock When the cookie meets the blockchain: Privacy risks of web payments
  via cryptocurrencies.
\newblock {\em PoPETs}, 2018:179--199, 2018.

\bibitem{revil}
A.~Goldsmith.
\newblock {What do we know about REvil, the Russian ransomware gang likely
  behind the Medibank cyber attack?}, November 2022.
\newblock
  \url{https://theconversation.com/what-do-we-know-about-revil-the-russian-ransomware-gang-likely-behind-the-medibank-cyber-attack-194337}.

\bibitem{wyb}
G.~Gomez, P.~Moreno-Sanchez, and J.~Caballero.
\newblock Watch your back: Identifying cybercrime financial relationships in
  bitcoin through back-and-forth exploration.
\newblock In {\em ACM SIGSAC Conference on Computer and Communications
  Security}, 2022.

\bibitem{gomez2023cybercrimeTR}
G.~Gomez, K.~van Liebergen, and J.~Caballero.
\newblock {Cybercrime Bitcoin Revenue Estimations: Quantifying the Impact of
  Methodology and Coverage}, 2023.
\newblock \url{https://arxiv.org/abs/2309.03592}.

\bibitem{harrigan2016unreasonable}
M.~Harrigan and C.~Fretter.
\newblock {The Unreasonable Effectiveness of Address Clustering}.
\newblock In {\em IEEE International Conference on Ubiquitous Intelligence and
  Computing (ATC)}, 2016.

\bibitem{graphsense}
B.~Haslhofer, R.~Stütz, M.~Romiti, and R.~King.
\newblock Graphsense: A general-purpose cryptoasset analytics platform.
\newblock {\em Arxiv pre-print}, 2021.
\newblock \url{https://arxiv.org/abs/2102.13613}.

\bibitem{hong2018you}
G.~Hong, Z.~Yang, S.~Yang, L.~Zhang, Y.~Nan, Z.~Zhang, M.~Yang, Y.~Zhang,
  Z.~Qian, and H.~Duan.
\newblock {How You Get Shot in the Back: A Systematical Study about
  Cryptojacking in the Real World}.
\newblock In {\em ACM SIGSAC Conference on Computer and Communications
  Security}, 2018.

\bibitem{trackingHuang}
D.~Y. Huang, M.~M. Aliapoulios, V.~G. Li, L.~Invernizzi, K.~McRoberts,
  E.~Bursztein, J.~Levin, K.~Levchenko, A.~C. Snoeren, and D.~McCoy.
\newblock {Tracking Ransomware End-to-end}.
\newblock In {\em IEEE Symposium on Security and Privacy}, May 2018.

\bibitem{huang2014botcoin}
D.~Y. Huang, H.~Dharmdasani, S.~Meiklejohn, V.~Dave, C.~Grier, D.~McCoy,
  S.~Savage, N.~Weaver, A.~C. Snoeren, and K.~Levchenko.
\newblock {Botcoin: Monetizing Stolen Cycles}.
\newblock In {\em Network and Distributed Systems Security Symposium}, 2014.

\bibitem{blocksci}
H.~Kalodner, M.~Möser, K.~Lee, S.~Goldfeder, M.~Plattner, A.~Chator, and
  A.~Narayanan.
\newblock {BlockSci: Design and Applications of a Blockchain Analysis
  Platform}.
\newblock In {\em USENIX Security Symposium}, 2020.

\bibitem{spamalytics}
C.~Kanich, C.~Kreibich, K.~Levchenko, B.~Enright, G.~M. Voelker, V.~Paxson, and
  S.~Savage.
\newblock Spamalytics: An empirical analysis of spam marketing conversion.
\newblock {\em Communications of the ACM}, page 99–107, sep 2009.

\bibitem{kappos22peel}
G.~Kappos, H.~Yousaf, R.~St{\"u}tz, S.~Rollet, B.~Haslhofer, and S.~Meiklejohn.
\newblock How to peel a million: Validating and expanding bitcoin clusters.
\newblock In {\em USENIX Security Symposium}, 2022.

\bibitem{cybercriminalLee}
S.~Lee, C.~Yoon, H.~Kang, Y.~Kim, Y.~Kim, D.~Han, S.~Son, and S.~Shin.
\newblock {Cybercriminal Minds: An Investigative Study of Cryptocurrency Abuses
  in the Dark Web}.
\newblock In {\em Network and Distributed Systems Security Symposium}, 2019.

\bibitem{li2023giveaway}
X.~Li, A.~Yepuri, and N.~Nikiforakis.
\newblock Double and nothing: Understanding and detecting cryptocurrency
  giveaway scams.
\newblock In {\em Network and Distributed Systems Security Symposium}, 2023.

\bibitem{behindLiao}
K.~Liao, Z.~Zhao, A.~Doup{\'e}, and G.-J. Ahn.
\newblock {Behind Closed Doors: Measurement and Analysis of CryptoLocker
  Ransoms in Bitcoin}.
\newblock In {\em APWG Symposium on Electronic Crime Research}, June 2016.

\bibitem{coinjoinMaxwell}
G.~Maxwell.
\newblock Coinjoin: Bitcoin privacy for the real world, August 2013.
\newblock \url{https://bitcointalk.org/index.php?topic=279249.0}.

\bibitem{fistfulMeiklejohn}
S.~Meiklejohn, M.~Pomarole, G.~Jordan, K.~Levchenko, D.~McCoy, G.~M. Voelker,
  and S.~Savage.
\newblock {A Fistful of Bitcoins: Characterizing Payments among Men with No
  Names}.
\newblock In {\em Internet Measurement Conference}, 2013.

\bibitem{raas}
P.~H. Meland, Y.~F.~F. Bayoumy, and G.~Sindre.
\newblock The ransomware-as-a-service economy within the darknet.
\newblock {\em Computers \& Security}, 92:101762, 2020.

\bibitem{inquiryMoser}
M.~{Möser}, R.~{Böhme}, and D.~{Breuker}.
\newblock {An Inquiry into Money Laundering Tools in the Bitcoin Ecosystem}.
\newblock In {\em APWG eCrime Researchers Summit}, September 2013.

\bibitem{bitcoin}
S.~Nakamoto.
\newblock Bitcoin: A peer-to-peer electronic cash system.
\newblock https://bitcoin.org/bitcoin.pdf, 2008.

\bibitem{nick2015msc}
J.~D. Nick.
\newblock {Data-Driven De-Anonymization in Bitcoin}.
\newblock Master's thesis, Distributed Computing Group, Computer Engineering
  and Networks Laboratory, ETH Zurich, Zurich, Switzerland, August 2015.

\bibitem{darksideDoS}
U.~D. of~State.
\newblock {DarkSide Ransomware as a Service (RaaS)}, November 2021.
\newblock \url{https://www.state.gov/darkside-ransomware-as-a-service-raas}.

\bibitem{omni}
Omni layer, 2023.
\newblock \url{https://www.omnilayer.org}.

\bibitem{oosthoek2022tale}
K.~Oosthoek, J.~Cable, and G.~Smaragdakis.
\newblock A tale of two markets: Investigating the ransomware payments economy,
  2022.

\bibitem{ransomwareClouston}
M.~Paquet-Clouston, B.~Haslhofer, and B.~Dupont.
\newblock {Ransomware Payments in the Bitcoin Ecosystem}.
\newblock {\em Journal of Cybersecurity}, 5(1), 2019.

\bibitem{spamsPaquetClouston}
M.~Paquet-Clouston, M.~Romiti, B.~Haslhofer, and T.~Charvat.
\newblock {Spams Meet Cryptocurrencies: Sextortion in the Bitcoin Ecosystem}.
\newblock In {\em ACM Conference on Advances in Financial Technologies}, 2019.

\bibitem{cerber}
S.~Pletinckx, C.~Trap, and C.~Doerr.
\newblock {Malware Coordination using the Blockchain: An Analysis of the Cerber
  Ransomware}.
\newblock In {\em IEEE Conference on Communications and Network Security},
  2018.

\bibitem{backpagePortnoff}
R.~S. Portnoff, D.~Y. Huang, P.~Doerfler, S.~Afroz, and D.~McCoy.
\newblock {Backpage and Bitcoin: Uncovering Human Traffickers}.
\newblock In {\em ACM SIGKDD International Conference on Knowledge Discovery
  and Data Mining}, 2017.

\bibitem{darkside}
T.~M. Research.
\newblock {What We Know About the DarkSide Ransomware and the US Pipeline
  Attack}, May 2021.
\newblock
  \url{https://www.trendmicro.com/en_us/research/21/e/what-we-know-about-darkside-ransomware-and-the-us-pipeline-attac.html}.

\bibitem{deadboltKeys}
Responders.NU, October 2022.
\newblock \url{https://deadbolt.responders.nu/}.

\bibitem{quantitativeRon}
D.~Ron and A.~Shamir.
\newblock {Quantitative Analysis of the Full Bitcoin Transaction Graph}.
\newblock In {\em Financial Cryptography and Data Security}, 2013.

\bibitem{dreadRon}
D.~Ron and A.~Shamir.
\newblock {How Did Dread Pirate Roberts Acquire and Protect his Bitcoin
  Wealth?}
\newblock In {\em Financial Cryptography and Data Security}, 2014.

\bibitem{coinshuffle}
T.~Ruffing, P.~Moreno-Sanchez, and A.~Kate.
\newblock Coinshuffle: Practical decentralized coin mixing for bitcoin.
\newblock In {\em Computer Security - ESORICS 2014}, pages 345--364. Springer
  International Publishing, 2014.

\bibitem{avclass}
M.~Sebastian, R.~Rivera, P.~Kotzias, and J.~Caballero.
\newblock {AVClass: A Tool for Massive Malware Labeling}.
\newblock In {\em Research in Attacks, Intrusions, and Defenses}, 2016.

\bibitem{shodan}
{Shodan}, 2022.
\newblock \url{https://www.shodan.io/}.

\bibitem{bitiodineSpagnoulo}
M.~Spagnuolo, F.~Maggi, and S.~Zanero.
\newblock {BitIodine: Extracting Intelligence from the Bitcoin Network}.
\newblock In {\em Financial Cryptography and Data Security}, 2014.

\bibitem{pony}
T.~Taniguchi, H.~Griffioen, and C.~Doerr.
\newblock {Analysis and Takeover of the Bitcoin-Coordinated Pony Malware}.
\newblock In {\em ACM ASIA Conference on Computer and Communications Security},
  2021.

\bibitem{conti}
F.~I. Team.
\newblock {Conti Ransomware: The History Behind One of the World’s Most
  Aggressive RaaS Groups}, October 2022.
\newblock \url{https://flashpoint.io/blog/history-of-conti-ransomware}.

\bibitem{tekiner2021sok}
E.~Tekiner, A.~Acar, A.~S. Uluagac, E.~Kirda, and A.~A. Selcuk.
\newblock {SoK: Cryptojacking Malware}.
\newblock In {\em IEEE European Symposium on Security and Privacy}, 2021.

\bibitem{thomas2015framing}
K.~Thomas, D.~Huang, D.~Wang, E.~Bursztein, C.~Grier, T.~J. Holt, C.~Kruegel,
  D.~McCoy, S.~Savage, and G.~Vigna.
\newblock Framing dependencies introduced by underground commoditization.
\newblock In {\em Workshop on the Economics of Information Security}, 2015.

\bibitem{laarschot2021risky}
J.~van~de Laarschot and R.~van Wegberg.
\newblock Risky business? investigating the security practices of vendors on an
  online anonymous market using {Ground-Truth} data.
\newblock In {\em 30th USENIX Security Symposium (USENIX Security 21)}, pages
  4079--4095. USENIX Association, Aug. 2021.
\newblock
  \url{https://www.usenix.org/conference/usenixsecurity21/presentation/van-de-laarschot}.

\bibitem{towardswu}
L.~Wu, Y.~Hu, Y.~Zhou, H.~Wang, X.~Luo, Z.~Wang, F.~Zhang, and K.~Ren.
\newblock Towards understanding and demystifying bitcoin mixing services.
\newblock In {\em Proceedings of the Web Conference 2021}, WWW '21, page
  33–44. Association for Computing Machinery, 2021.

\bibitem{xia2020characterizing}
P.~Xia, H.~Wang, B.~Zhang, R.~Ji, B.~Gao, L.~Wu, X.~Luo, and G.~Xu.
\newblock {Characterizing Cryptocurrency Exchange Scams}.
\newblock {\em Computers \& Security}, 98, 2020.

\end{thebibliography}

	%-- Bib style for citations of the form: [Zippy97]
	%\bibliographystyle{$REFDIR/tex/macros/refalpha}

	%-- Citations by number: [23]
	%\bibliographystyle{$REFDIR/tex/macros/refplain}

	%-- Citations by number w/ only the authors' 1st and middle initials
	%\bibliographystyle{$REFDIR/tex/macros/refabbrv}
}

\appendix

% \section{Appendix}
% \label{sec:appendix}

\begin{table*}
	\begin{minipage}{0.49\linewidth}
		\scriptsize
	\scalebox{0.81}{
	\begin{tabular}{l|r|r|r|r|r|r|r}
	\hline
	\textbf{Family} & \textbf{Seeds} & \textbf{OW} & \textbf{Clust} & \textbf{Addr.} &\textbf{Activity Period} & \textbf{BTC} & \textbf{USD} \\
	\hline
	R:Netwalker	& 66	& 0	& 65	& 329	& 20180216:20210528	& 3,131.7562	& \$28,262,120 \\ 	R:Conti	& 26	& 0	& 26	& 35	& 20171204:20220207	& 410.8025	& \$18,869,852 \\
	R:REvil	& 7	& 0	& 6	& 77	& 20190823:20210611	& 369.5363	& \$12,540,689 \\ 	R:DarkSide	& 3	& 0	& 3	& 3	& 20210215:20210511	& 158.7086	& \$9,124,091 \\
	R:Locky	& 7,036	& 0	& 2	& 7,094	& 20160114:20170602	& 15,396.6934	& \$7,833,613 \\
	R:Ryuk	& 26	& 0	& 26	& 39	& 20180811:20230322	& 866.8069	& \$4,844,149 \\
	R:RagnarLocker	& 1	& 0	& 1	& 1	& 20200727:20200728	& 414.0007	& \$4,547,241 \\
	R:MountLocker	& 1	& 0	& 1	& 1	& 20201103:20201104	& 298.5000	& \$4,226,277 \\
	R:BlackMatter	& 1	& 0	& 1	& 1	& 20210730:20210730	& 96.3863	& \$4,068,295 \\
	R:Makop	& 1	& 0	& 1	& 996	& 20190617:20230408	& 316.0883	& \$3,853,713 \\
	R:Egregor	& 9	& 0	& 9	& 9	& 20201023:20201207	& 197.9554	& \$3,129,599 \\
	R:Tejodes	& 1	& 0	& 1	& 178	& 20171117:20200928	& 345.5527	& \$2,798,239 \\
	P:Leancy	& 1	& 0	& 1	& 1	& 20131216:20171123	& 3,092.8482	& \$1,896,583 \\
	R:CryptXXX	& 1	& 0	& 1	& 1,742	& 20160511:20161006	& 3,338.5464	& \$1,877,833 \\
	R:DMALockerv3	& 9	& 0	& 3	& 177	& 20160818:20211105	& 1,833.9286	& \$1,757,464 \\
	R:MedusaLocker	& 3	& 0	& 3	& 26	& 20200603:20220121	& 56.1546	& \$1,454,776 \\
	R:HelloKitty	& 1	& 0	& 1	& 1	& 20210712:20210712	& 32.3529	& \$1,070,653 \\
	R:Bitpaymer	& 1	& 0	& 1	& 1	& 20200813:20200817	& 90.0004	& \$1,058,347 \\ 	C:cliptomaner	& 1	& 1	& 1	& 1	& 20191117:20230411	& 38.7643	& \$1,004,397 \\
	P:Cryptory	& 1	& 1	& 1	& 1	& 20130902:20150108	& 1,677.5167	& \$886,689 \\
	R:Qlocker	& 22	& 0	& 4	& 45	& 20210420:20220911	& 14.9621	& \$786,335 \\
	R:Spora	& 1	& 0	& 1	& 2,129	& 20170105:20210102	& 616.2996	& \$723,352 \\
	R:SamSam	& 23	& 0	& 21	& 51	& 20160113:20170630	& 646.2379	& \$583,508 \\
	R:CryptoLocker	& 2	& 0	& 1	& 968	& 20130907:20170710	& 1,511.7579	& \$520,238 \\
	C:clipsa	& 576	& 0	& 37	& 784	& 20180821:20230324	& 25.9030	& \$507,931 \\
	P:RockwellPartners	& 1	& 0	& 1	& 1	& 20140303:20150526	& 731.7512	& \$383,136 \\
	P:bitcoindoubler.fund	& 1	& 1	& 1	& 1	& 20170803:20180127	& 49.5862	& \$323,677 \\
	P:MiniPonziCoin	& 1	& 0	& 1	& 1	& 20140218:20141022	& 445.9718	& \$266,979 \\
	P:Ponzi.io	& 2	& 0	& 1	& 33	& 20140204:20171030	& 369.3729	& \$258,604 \\
	P:BTC-doubler.com	& 1	& 0	& 1	& 1	& 20161226:20170723	& 176.3050	& \$229,916 \\
	R:SunCrypt	& 1	& 0	& 1	& 1	& 20210719:20210719	& 7.4484	& \$229,694 \\
	C:clipbanker	& 4	& 1	& 2	& 186	& 20160422:20230412	& 17.3412	& \$210,100 \\
	C:phorpiextldr	& 15	& 14	& 2	& 18	& 20180531:20230302	& 19.8600	& \$198,675 \\
	C:slave	& 1	& 0	& 1	& 1	& 20150102:20170714	& 651.4415	& \$193,837 \\
	C:n40	& 1	& 1	& 1	& 1	& 20170531:20191002	& 27.4139	& \$187,461 \\
	R:AES-NI	& 1	& 0	& 1	& 1	& 20140606:20210506	& 85.6786	& \$183,968 \\
	R:NoobCrypt	& 1	& 0	& 1	& 28	& 20131018:20160715	& 550.4330	& \$164,735 \\
	R:LockBit 2.0	& 2	& 0	& 2	& 51	& 20200929:20211008	& 4.9379	& \$158,709 \\
	P:Nanoindustryinv.com	& 1	& 0	& 1	& 2	& 20140816:20150213	& 480.1777	& \$155,317 \\
	C:mrpr0gr4mmer	& 3	& 0	& 2	& 377	& 20191226:20230410	& 9.0049	& \$130,166 \\
	R:WannaCry	& 5	& 0	& 4	& 6	& 20170331:20230306	& 59.5107	& \$119,094 \\
	R:SynAck	& 1	& 0	& 1	& 1	& 20150329:20180122	& 99.2369	& \$114,116 \\
	R:GlobeImposter	& 1	& 0	& 1	& 1	& 20141123:20171227	& 190.1388	& \$97,266 \\
	P:bitcoincopy.site123.me	& 1	& 1	& 1	& 1	& 20160725:20181112	& 17.9770	& \$91,826 \\
	P:1hourbtc.pw	& 1	& 0	& 1	& 433	& 20161031:20210104	& 43.4324	& \$89,381 \\
	R:AvosLocker	& 1	& 0	& 1	& 1	& 20210909:20210909	& 1.8047	& \$83,722 \\
	P:10PERCENTBTC	& 1	& 0	& 1	& 17	& 20140812:20230403	& 115.1086	& \$81,443 \\
	P:Ponzi120	& 1	& 0	& 1	& 1	& 20140225:20140301	& 144.9008	& \$77,763 \\
	C:masad	& 2	& 0	& 2	& 114	& 20170117:20230227	& 5.6161	& \$76,321 \\
	C:cryptoshuffler	& 1	& 0	& 1	& 156	& 20160303:20210831	& 65.1258	& \$71,890 \\
	C:phorpiextrik	& 7	& 1	& 7	& 9	& 20160814:20230104	& 11.4488	& \$68,726 \\
	P:GrandAgoFinance	& 1	& 0	& 1	& 1	& 20140924:20150129	& 178.7435	& \$64,765 \\
	R:Ako	& 1	& 0	& 1	& 14	& 20200107:20210530	& 5.9315	& \$54,487 \\
	P:investorbitcoin.com	& 1	& 1	& 1	& 1	& 20160531:20171003	& 47.3833	& \$51,444 \\
	P:1getpaid.me	& 1	& 0	& 1	& 2	& 20150607:20160429	& 176.7322	& \$43,250 \\
	R:APT	& 2	& 0	& 1	& 3	& 20160915:20171130	& 30.3040	& \$39,348 \\
	R:Globev3	& 5	& 0	& 5	& 19	& 20160925:20180220	& 40.4265	& \$36,493 \\
	P:LaxoTrade	& 1	& 1	& 1	& 1	& 20140820:20181218	& 69.7857	& \$25,998 \\
	R:Flyper	& 2	& 0	& 1	& 31	& 20161004:20181013	& 7.4181	& \$25,011 \\
	C:mekotion40	& 3	& 0	& 2	& 20	& 20180125:20221013	& 2.4035	& \$24,950 \\
	P:Twelverized	& 1	& 0	& 1	& 1	& 20140302:20140413	& 38.8018	& \$24,599 \\
	R:Globe	& 3	& 0	& 3	& 87	& 20160605:20170115	& 31.9565	& \$23,577 \\
	C:predatorthethief	& 1	& 0	& 1	& 39	& 20190809:20201115	& 2.4541	& \$23,461 \\
	C:mekotio	& 2	& 0	& 2	& 14	& 20170712:20220530	& 1.5535	& \$22,189 \\
	C:azorult	& 1	& 1	& 1	& 1	& 20180719:20190324	& 5.9413	& \$21,872 \\
	R:LockBit	& 1	& 0	& 1	& 1	& 20201201:20201214	& 1.1114	& \$21,007 \\
	C:clipboardwallethijacker	& 4	& 0	& 1	& 34	& 20180302:20220318	& 2.6740	& \$18,721 \\
	C:mispadu	& 1	& 1	& 1	& 1	& 20190928:20220927	& 0.7798	& \$18,473 \\
	R:EDA2	& 2	& 0	& 2	& 33	& 20140406:20170725	& 10.8830	& \$18,299 \\
	P:bestdoubler.eu	& 1	& 0	& 1	& 8	& 20161223:20170831	& 13.2392	& \$18,149 \\
	R:Sam	& 1	& 0	& 1	& 1	& 20160927:20160927	& 29.0000	& \$17,538 \\
	\hline
	\end{tabular}
	}
							\end{minipage}
		\begin{minipage}{0.49\linewidth}

			\scriptsize
	\scalebox{0.81}{
	\begin{tabular}{l|r|r|r|r|r|r|r}
	\hline
	\textbf{Family} & \textbf{Seeds} & \textbf{OW} & \textbf{Clust} & \textbf{Addr.} &\textbf{Activity Period} & \textbf{BTC} & \textbf{USD} \\
	\hline
	R:CryptoTorLocker2015	& 7	& 0	& 1	& 159	& 20131106:20170312	& 52.7342	& \$15,068 \\
	R:Black Kingdom	& 2	& 0	& 2	& 2	& 20200306:20210318	& 0.7233	& \$15,015 \\
	R:NotPetya	& 3	& 0	& 3	& 3	& 20170626:20230306	& 4.8104	& \$13,770 \\
	C:casbaneiro	& 1	& 0	& 1	& 19	& 20190307:20211109	& 1.6425	& \$13,691 \\
	R:XLockerv5.0	& 2	& 0	& 2	& 3	& 20170124:20170307	& 12.0000	& \$12,568 \\
	P:BTC-doubler.us	& 1	& 0	& 1	& 7	& 20170502:20180512	& 4.3549	& \$12,527 \\
	R:Xorist	& 3	& 1	& 3	& 11	& 20170523:20171001	& 4.3138	& \$11,850 \\
	R:File-Locker	& 1	& 1	& 1	& 1	& 20120309:20230322	& 2.8771	& \$10,832 \\
	C:kryptocibule	& 8	& 0	& 4	& 13	& 20200228:20230308	& 0.7980	& \$10,706 \\
	R:XTPLocker	& 1	& 0	& 1	& 4	& 20160509:20161110	& 19.1398	& \$10,283 \\
	R:StorageCrypter	& 1	& 0	& 1	& 5	& 20171110:20171208	& 0.9668	& \$9,908 \\
	P:bitcoindoubler.prv.pl 	& 1	& 0	& 1	& 1	& 20160516:20220217	& 11.5914	& \$9,758 \\
	R:Ranzy Locker	& 1	& 0	& 1	& 1	& 20210226:20210226	& 0.2100	& \$9,729 \\
	R:Ransomnix	& 1	& 0	& 1	& 1	& 20171121:20180704	& 1.2496	& \$9,474 \\
	P:120cycle	& 1	& 0	& 1	& 78	& 20140324:20140329	& 14.2823	& \$8,264 \\
	P:Minimalism10	& 1	& 0	& 1	& 1	& 20140227:20140327	& 13.0614	& \$7,839 \\
	P:CRYPTOSX2	& 1	& 0	& 1	& 52	& 20150129:20180106	& 27.2141	& \$7,490 \\
	R:LamdaLocker	& 1	& 1	& 1	& 1	& 20170110:20170731	& 6.6692	& \$7,430 \\
	R:ChupaCabra	& 1	& 1	& 1	& 1	& 20210513:20210923	& 0.1907	& \$7,362 \\
	R:AlbDecryptor	& 1	& 0	& 1	& 18	& 20200423:20211118	& 0.2836	& \$7,355 \\
	C:protonbot	& 2	& 2	& 2	& 2	& 20180417:20210409	& 0.9430	& \$7,349 \\
	R:Vega	& 1	& 0	& 1	& 11	& 20210809:20210920	& 0.1463	& \$6,989 \\ 	P:doublebitcoin.life	& 1	& 1	& 1	& 1	& 20151026:20201016	& 3.9236	& \$6,853 \\
	R:VenusLocker	& 2	& 0	& 2	& 5	& 20160719:20170217	& 6.8115	& \$6,742 \\
	R:Phobos	& 1	& 1	& 1	& 1	& 20210720:20210720	& 0.1845	& \$5,497 \\
	R:CryptConsole	& 6	& 0	& 6	& 7	& 20160428:20170914	& 3.8135	& \$4,254 \\
	R:Predator	& 1	& 1	& 1	& 1	& 20180502:20190629	& 0.5559	& \$3,522 \\
	R:JigSaw	& 6	& 0	& 5	& 20	& 20160328:20170315	& 5.2538	& \$3,369 \\
	R:Qweuirtksd	& 1	& 0	& 1	& 11	& 20181025:20181117	& 0.5407	& \$3,301 \\
	R:Black Ruby	& 1	& 0	& 1	& 1	& 20180205:20180213	& 0.3678	& \$3,054 \\
	R:Git	& 1	& 0	& 1	& 4	& 20190503:20211114	& 0.3497	& \$3,006 \\
	R:XLocker	& 1	& 0	& 1	& 1	& 20170524:20170524	& 1.0000	& \$2,476 \\
	C:bitcoingrabber	& 1	& 0	& 1	& 16	& 20190130:20220807	& 0.2821	& \$2,443 \\
	P:igjam.com	& 1	& 0	& 1	& 2	& 20131211:20170616	& 3.0101	& \$2,291 \\
	R:HC6/HC7	& 6	& 0	& 6	& 6	& 20170729:20171127	& 0.3206	& \$2,121 \\
	P:world-btc.online	& 1	& 0	& 1	& 51	& 20170120:20170924	& 1.9167	& \$2,082 \\
	R:Bagli	& 1	& 0	& 1	& 8	& 20210409:20230403	& 0.0825	& \$1,971 \\
	R:Cryptowall	& 1	& 0	& 1	& 4	& 20160429:20161116	& 2.6277	& \$1,689 \\
	P:OpenPonzi	& 1	& 0	& 1	& 1	& 20150415:20160308	& 6.0855	& \$1,490 \\
	R:ComradeCircle	& 1	& 0	& 1	& 1	& 20161018:20161018	& 2.0332	& \$1,292 \\
	R:Avaddon	& 1	& 0	& 1	& 1	& 20210512:20210512	& 0.0241	& \$1,193 \\
	R:Cryptohitman	& 2	& 0	& 1	& 15	& 20160427:20190923	& 1.9001	& \$1,049 \\
	R:Encrpt3d	& 1	& 0	& 1	& 1	& 20191209:20210704	& 0.1107	& \$1,015 \\
	C:androidclipper	& 1	& 1	& 1	& 1	& 20180223:20181206	& 0.1287	& \$942 \\
	R:Ecovector	& 1	& 0	& 1	& 2	& 20160511:20160521	& 2.0041	& \$888 \\
	R:Decryptiomega	& 2	& 1	& 2	& 2	& 20190718:20190807	& 0.0600	& \$705 \\
	P:bitcoinprofit2	& 1	& 0	& 1	& 1	& 20160221:20200301	& 0.5118	& \$672 \\
	R:Kelly	& 1	& 0	& 1	& 1	& 20200909:20200909	& 0.0500	& \$502 \\
	R:Gula	& 1	& 1	& 1	& 1	& 20180919:20181117	& 0.0425	& \$268 \\
	R:DMALocker	& 1	& 0	& 1	& 1	& 20151228:20151228	& 0.6000	& \$253 \\
	P:7dayponzi	& 1	& 0	& 1	& 1	& 20150321:20150526	& 0.7735	& \$202 \\
	R:Vevolocker	& 1	& 0	& 1	& 5	& 20170923:20171116	& 0.0227	& \$152 \\
	C:aggah	& 1	& 0	& 1	& 1	& 20200518:20200826	& 0.0130	& \$141 \\
	R:Black Mamba	& 1	& 1	& 1	& 1	& 20170619:20170619	& 0.0399	& \$105 \\
	P:invest4profit	& 1	& 0	& 1	& 11	& 20161029:20170929	& 0.0317	& \$42 \\
	R:LockOn	& 1	& 1	& 1	& 1	& 20170826:20170826	& 0.0050	& \$22 \\
	R:BlackRouter	& 1	& 1	& 1	& 1	& 20181219:20190528	& 0.0023	& \$15 \\
	R:NullByte	& 1	& 0	& 1	& 1	& 20160903:20160903	& 0.0172	& \$10 \\
	R:Exotic	& 1	& 0	& 1	& 1	& 20161028:20161028	& 0.0064	& \$4 \\
	R:Bucbi	& 1	& 0	& 1	& 1	& 20160402:20160402	& 0.0090	& \$4 \\
	R:PopCornTime	& 1	& 1	& 1	& 1	& 20161214:20161230	& 0.0028	& \$2 \\
	R:Phoenix	& 1	& 0	& 1	& 1	& 20161214:20161214	& 0.0026	& \$2 \\
	R:CryptoHost	& 1	& 0	& 1	& 1	& 20160612:20160612	& 0.0020	& \$1 \\
	R:WannaRen	& 1	& 0	& 1	& 1	& 20200408:20200408	& 0.0001	& \$1 \\
	R:7ev3n	& 1	& 0	& 1	& 1	& 20160429:20160429	& 0.0011	& \$1 \\
	R:Chimera	& 1	& 0	& 1	& 1	& 20151107:20151108	& 0.0005	& \$0 \\
	R:WannaSmile	& 1	& 0	& 1	& 1	& 20171110:20171110	& 0.0000	& \$0 \\
	R:CTB-Locker	& 1	& 0	& 1	& 1	& 20160313:20160418	& 0.0002	& \$0 \\
	R:TeslaCrypt	& 1	& 0	& 1	& 1	& 20150315:20150315	& 0.0001	& \$0 \\
	P:btcgains	& 1	& 1	& 1	& 1	& -	& 0.0000	& \$0 \\
	\hline
	Total	& 8,021	& 42	& 383	& 17,040	& 20120309:20230412	& 40,282.3312	& \$123,357,041 \\
	\hline
	\end{tabular}
	}
					
	\end{minipage}
	\caption{Operations in the Ransomwhere~\cite{ransomwhere},
	Ponzi~\cite{ponziBartoletti}, and Clipper~\cite{wyb}
	datasets, ordered by USD revenue using the \emph{DD-OW+MI-DC} estimation. 
	The last row captures total revenue of all 141 operations.}
	\label{tab:ransomwhere_full}

\end{table*}

\end{document}